\documentclass[pdftex, conference]{IEEEtran}
\usepackage{float}
\usepackage{caption}
\usepackage{subfig}
\usepackage{tabularx}
\usepackage{amsmath}
\usepackage{graphicx}
\usepackage{amssymb}
\graphicspath{{./figures/ }}
\usepackage[rightcaption]{sidecap}
\usepackage{wrapfig}
\usepackage{xcolor}
\usepackage{multirow}
\usepackage{url}
\usepackage[normalem]{ulem}
\usepackage[english]{babel}
\usepackage{paralist, tabularx}
\usepackage{comment}
\usepackage{authblk}
\usepackage{pifont}
\def\fplus{\texttt{+}}
\def\fminus{\texttt{-}}

\newcommand{\cmark}{\ding{51}}%
\newcommand{\xmark}{\ding{55}}%
\newcolumntype{H}{>{\setbox0=\hbox\bgroup}c<{\egroup}@{}}
\newcommand{\ignore}[1]{\if{0} #1 \fi}
\newcounter{inlineenum}
\renewcommand{\theinlineenum}{\alph{inlineenum}}

\newcounter{subsubsubsection}[subsubsection]

\begin{document}
\title{SoK: The Faults in our ASRs: An Overview of Attacks against Automatic Speech Recognition and Speaker Identification Systems}
\author[1]{Hadi Abdullah}
\author[1]{Kevin Warren}
\author[1]{Vincent Bindschaedler}
\author[2]{Nicolas Papernot}
\author[1]{Patrick Traynor}
\affil[1]{University of Florida}
\affil[2]{University of Toronto}

\maketitle




\begin{abstract}

Speech and speaker recognition systems are employed in a variety of
    applications, from personal assistants to telephony surveillance and
    biometric authentication.  The wide deployment of these systems has
    been made possible by the improved accuracy in neural networks.  Like other systems based on
    neural networks, recent research has demonstrated that speech and
    speaker recognition systems are vulnerable to attacks using
    manipulated inputs. However, as we demonstrate in this paper, the
    end-to-end architecture of speech and speaker systems and the nature
    of their inputs make attacks and defenses against them
    substantially different than those in the image space. We
    demonstrate this first by systematizing existing research in this space and providing a taxonomy through which the
    community can evaluate future work. We then demonstrate
    experimentally that attacks against these models almost universally
    fail to transfer. In so doing, we argue that
    substantial additional work is required to provide adequate
    mitigations in this space.

\end{abstract}

\section{Introduction}

Voice Processing Systems (VPSes) are a critical interface for both
classical and emerging systems. While by no means conceptually new, the
last decade has seen a dramatic improvement in both Automatic Speech
Recognition (ASR) and Speaker Identification (SI) VPSes. Such interfaces
are not merely for convenience; rather, they drastically improve
usablity for groups such as the elderly and the visually
impared~\cite{CM17, HS18}, make devices without screens such as headless
IoT systems accessible~\cite{article_acc_models}, and make user
authentication nearly invisible~\cite{azure_attest, mozilla_ds}.

Advances in neural networks have helped make VPSes practical.
Although different architectures have been used in the past
(e.g., Hidden Markov models), systems built atop neural networks now
dominate the space. While neural networks have enabled significant
improvements in transcription and identification accuracy, substantial
literature in the field of adversarial machine learning shows that they
are also vulnerable to a wide array of attacks. In particular, the
research community has put forth significant effort to demonstrate that
image classification and, only recently, VPSes built on neural
networks are vulnerable to exploitation using small perturbations to
their inputs. 

While it may be tempting to view VPSes as simply another application of
neural networks and to therefore assume that previous work on
adversarial machine learning applies directly to this new application,
this paper shows that this is demonstrably untrue. We
make the following contributions:

\begin{compactitem}

    \item {\bf Taxonomization of VPS Threat Models:} Attacks on VPSes
	are conducted using a number of widely differing (sometimes implicit) assumptions
	about adversarial behavior and ability.
	We provide the first framework for reasoning more broadly about
	work in this space.

    \item {\bf Categorization of Existing Work:} We take the body of
	work on attacks and defenses for VPSes and categorize them
	based on the above taxonomy. We show that while many papers have
	already been published, significant work remains to be done.

    \item {\bf Experimental Testing of Transferability:} We demonstrate
	that \emph{transferability}, or the ability to exploit multiple models using 
	an adversarial input is not
	currently achievable against VPSes via attacks that rely on gradient-based optimization~\cite{carlini2018audio}. Through
	extensive experimentation, we show that transferability
	is currently \emph{extremely} unlikely even when considering two
	instances of the same VPSes trained separately on the same train-test splits,
	hyper-parameters, initial random seeds and architecture. The methodology and results for the experiments can be found in the Appendix.

\end{compactitem}


The remainder of this paper is organized as follows:
Section~\ref{background} provides background information on VPSes;
Section~\ref{signal_processing_attack} discusses the special
considerations that must be understood when attacking VPSes;
Sections~\ref{def_threat_model} and~\ref{attack_threat_model} details our threat model taxonomy;
Section~\ref{overview} then identifies the novel contributions of
published work through this taxonomy; Section~\ref{defense} explores
currently proposed defense and detection mechanisms;
Section~\ref{transferability_expts} explores transferability and why
optimization attacks currently fail to provide this property for VPSes;
Section~\ref{discussion} discusses open issues; Section~\ref{conc}
provides concluding remarks and the Appendix details the methodology and results for the transferability experiments.

\section{Background}\label{background}
\begin{figure*}
  \includegraphics[width=\linewidth]{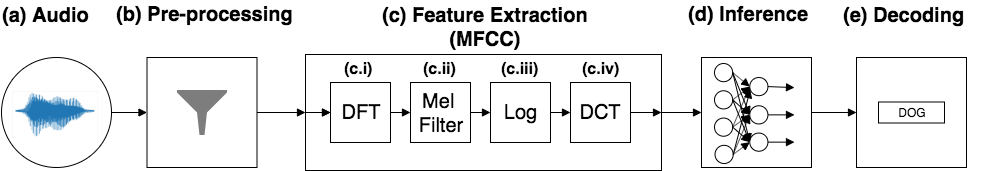}
  \caption{The steps that form the Automatic Speech Recognition
pipeline. (a) The audio is recorded using a microphone. (b) It is
prepossessed to remove any rudimentary noise or high frequencies using
a low pass filter. (c) The audio is passed through a feature extraction
function (in this case the MFCC function) that extracts the most important
features of the audio sample. (d) The features are passed on to the
model for inference, which outputs a non-human-readable string. (e) The
string is decoded to produce a human-readable transcription.}
\label{fig:pipeline} 
\vspace{-0.8em}
\hrulefill
 \vspace{-2em}
\end{figure*}
\subsection{Psychoacoustics}\label{acoustics}\label{perception_mechanism}\label{hearing_range}\label{auditory_masking}\label{cocktail_party_effect}\label{non_linearity}

Modern speech recognition systems are designed to approximate the functioning
of the human auditory system~\cite{biometrics}. Research in
psychoacoustics has revealed the complex ways with which the human brain
processes audio. For example, there is a difference between the actual intensity and
the brain's perceived intensity of an audio tone. Therefore, the perception of loudness
is non-linear with respect to intensity~\cite{lin2015principles}.
In order to double the brain's perceived
intensity, one must increase the actual intensity by a factor of eight.
Similarly, the limit of human hearing ranges from
20Hz to 20kHz. Any frequency outside this range (e.g., ultrasound (20kHz to 10
MHz)), can not be heard. The brain employs complex processes to understand
speech in noisy environments or during cross-talk. These include visual cues, pitch
separation, intensity, binural unmasking, context, and
memory~\cite{lin2015principles}. 

\subsection{Voice Processing Systems (VPSes)}\label{asr_pipe}
Researchers have developed VPSes to capture functionality of the human auditory system, albeit
coarsely.  VPSes are of two types: ASR systems and SI systems. ASR
systems convert speech to text. SI systems attempt to identify a person by
their speech samples. There are a wide variety of ASR and SI systems in use today, including personal assistants
(e.g., Google Home~\cite{google}, Amazon Alexa~\cite{amazon},
Siri~\cite{siri}), telephony surveillance systems~\cite{NSA_1,NSA_2,NSA_3,NSA_4,NSA_5}, and
conferencing transcription services~\cite{otter}.


The modern ASR system pipeline consists of the steps shown in~Figure\ref{fig:pipeline}.
Due to the similarity of the SI and ASR pipelines, in this subsection
we define the pipeline in the context of ASR. The steps include the following:

\subsubsection{Preprocessing}
The audio sample is first passed through a preprocessing phase
(Figure~\ref{fig:pipeline}(b)). Here, segments of
audio containing human speech are identified using voice-activity detection algorithms, such
as G.729~\cite{sohn1999statistical}. Next, these selected segments are passed
through a low-pass filter to remove high frequencies. This helps to improve
ASR accuracy by removing the unnecessary noise from the samples. 

\subsubsection{Feature Extraction}
The filtered audio is divided into overlapping frames, usually 20 ms in
length~\cite{sigurdsson2006mel}, to capture the transitions in the
signal. Each frame is then passed to the feature extraction step
(Figure~\ref{fig:pipeline}(c)). To approximate the human auditory system, the
most commonly used algorithm is the Mel Frequency Cepstral Coefficient (MFCC)~\cite{sigurdsson2006mel},
which extracts the features that the human ear considers most important. The
MFCC is comprised of four steps (Figure~\ref{fig:pipeline}(c)):

\paragraph{Discrete Fourier Transform (DFT)}\label{dft}
The DFT of the audio sample is computed first
(Figure~\ref{fig:pipeline}(c.i)). It converts a time domain signal into
the frequency domain representation~\cite{rabiner1978digital}:

\begin{eqnarray*}
    F_k = \sum_{n=0}^{N-1}s_n \left(\mathrm{cos}[\frac{\pi}{N}(n+\frac{1}{2})k] 
    -i \cdot \mathrm{sin}[\frac{\pi}{N}(n+\frac{1}{2})k]\right) \\
  \nonumber 
\end{eqnarray*}
where the real and imaginary parts are used to infer the phase of the
signal and intensity of a frequency. $F = (F_1,...,F_N)$ is a vector of
complex numbers, $N$ is the total number of samples, $n$ is the sample
number, $k$ is the frequency number, and $s_n $ is the $n$-th sample of
the input. The magnitude of $F_k$ corresponds to the intensity of the
$k$-th frequency in $s$. The DFT output, called the spectrum,
provides a fine-grained understanding of an audio sample's frequency
composition.

\paragraph{Mel Filtering} \label{mel_filtering}
As previously mentioned, the human auditory
system treats frequencies in a non-linear manner. To recreate this
effect, the Mel filter scales the intensities of the frequencies
accordingly (Figure~\ref{fig:pipeline}(c.ii)), using the following:
\begin{eqnarray*}
m_k=2595\log _{10}\left(1+{\frac {|F_k|}{700}}\right)
\end{eqnarray*}
where $m_k$ is the resulting Mel scaled frequency intensity. The scale assigns a higher
weight to frequencies that exist between 100Hz and
8kHz~\cite{ahmed1974discrete}. This is done to amplify the frequencies
that constitute human speech.

\paragraph{Log Scaling}\label{logarithm_of_powers}
The output of the Mel-filter is scaled by the log function
(Figure~\ref{fig:pipeline}(c.iii)). This reproduces logarithmic
perception exhibited by the human auditory system.

\paragraph{Discrete Cosine Transform (DCT)}\label{dct}
The DCT~(Figure~\ref{fig:pipeline}(c.iv))~\cite{ahmed1974discrete} decomposes an input into a series of cosine components. The components that represent most of the information about the input are retained, while the rest are discarded. This is done using the following equation: 
\begin{eqnarray*}
F_k & = & \sum_{n=0}^{N-1} s_n \cos\left( \frac{ (2n+1)k\pi}{2N}\right) 
\end{eqnarray*}
where $F_k$ is the intensity of the $k^{th}$ component in $s$.

\subsubsection{Inference}\label{infer}
The extracted features are finally passed to a probabilistic model for
inference (Figure~\ref{fig:pipeline}(d)). There are a variety of models
available for VPSes. However, we focus on neural networks as they are the dominant choice in this space.


\paragraph{Convolutional Neural Network (CNN)}\label{dnn}\label{cnn}
A neural network with more than one hidden layers is called a Deep Neural Network (DNN). A special type of DNN is a CNN in which each layer of the CNN is made up of a set of filters that are convolved with the layer's input to obtain a new representation of the data. However, interpreting
exactly what each hidden layer has learned is an open
problem~\cite{gunning2017explainable}.

CNNs have one major limitation: they take a fixed-sized input
and produce a fixed-sized output. This is ideal for image
recognition, where images can be down-sampled or up-sampled to a
specific size. However CNNs may be too constrained for applications such
as speech recognition, where each input can be of arbitrary length and
down-sampling can result in loosing contextual information.
\begin{figure}
  \centering
  \includegraphics[width=\linewidth,height=5cm]{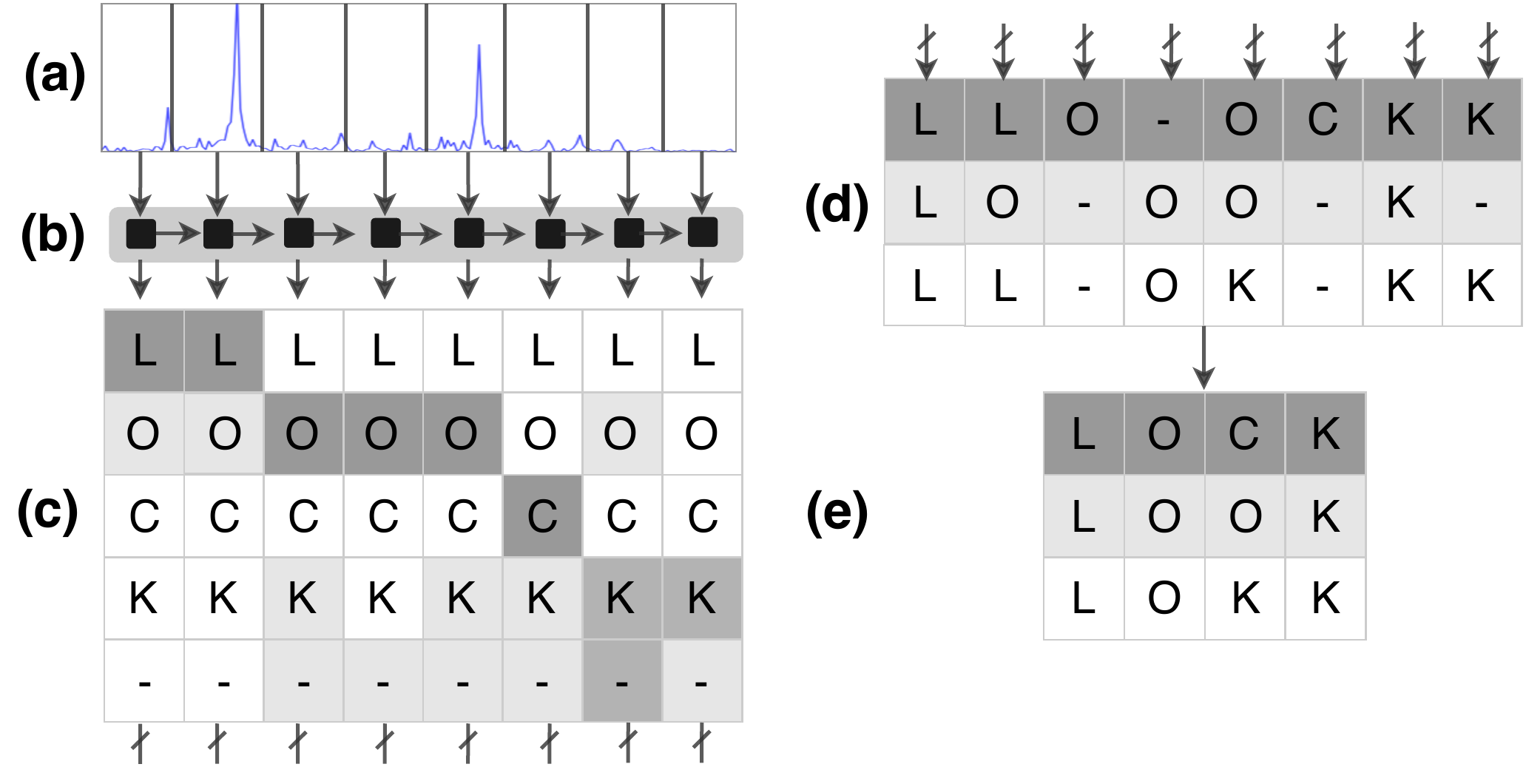}
  \caption{Example of a simple beam search decoder. The
input data (a) is fed into an RNN as shown in (b) which gives the
distribution of the outputs. In this example the possible output
characters are \{l,o,c,k,-\} shown in (c) and with their distributions
the beam search calculates the probability of different sequences (d).
Using those sequence probabilities, the CTC beam search calculates the
probability of different outputs. } 
\label{fig:ctc}
\vspace{-0.8em}
\hrulefill
\vspace{-2em}
\end{figure}

\paragraph{Recurrent Neural Network (RNN)}\label{rnn}
An RNN overcomes this limitation of
CNNs. An RNN takes a variable-sized input and can produce a
variable-sized output. This is ideal for VPSes, as
speech input can be of variable length as a result of varying
pronunciation speeds of the speaker and different sentences having a variable number of words. Additionally, RNNs are designed to
use contextual information, which is important for VPSes. 
An RNN accumulates this context in a hidden state, which is fed to the RNN as it processes the following steps of an input sequence. The hidden state acts as an
internal memory unit by `remembering' the information about the
previous time steps. For example, consider the phrase ``Mary had a
little $\rule{1cm}{0.15mm}$''. Intuitively, an RNN uses information
about the past to fill in the missing word (i.e., ``lamb''). This 
behavior is critical for ASRs as human
speech has a relatively consistent temporal structure: the word at time
$t_n$ generally depends on words spoken at $t_m$ where $m<n$. 

\subsubsection{Decoding}
The decoding stage is shown in Figure~\ref{fig:ctc}. 
For each 20 ms frame (Figure~\ref{fig:ctc}(a)), the inference step
(Figure~\ref{fig:ctc}(b)) produces a probability distribution over all
the characters (Figure~\ref{fig:ctc}(c)). The model then produces
a two-dimensional (character $\times$ total time) matrix of probabilities
(Figure~\ref{fig:ctc}(d)). In the image domain, one typically picks the
label corresponding to the largest probability as the final output. In contrast, doing so in an ASR
will not necessarily produce a correct transcription. As a result,
there are multiple possible transcriptions for a single
audio file~\cite{hannun2017sequence}. For example, as seen in
Figure~\ref{fig:ctc}(d), for the audio sample with the word ``lock'', a
model may produce multiple strings such as ``llo-ockk'' or ``ll-ok-kk''. The
output strings are not reader friendly, since they reflect speaker
characteristics like speed and accent. To overcome this problem,
the decoding stage of the ASR system converts the model output string
into words (Figure~\ref{fig:ctc}(e)). One of the most commonly used
decoding algorithms is Beam Search~\cite{koehn2004pharaoh}. It is charged with
selecting a sequence of tokens
based on a
distribution of probabilities over dictionary words,
which the model predicts for each token. This heuristic is commonly employed
to explore multiple sequences without being too sensitive to the model's prediction at 
each step of the sequence.  This is a
heuristic that always outputs the most likely word for a given label
string, thus the output transcription is not
guaranteed to be optimal.

\subsubsection{Alternative Configurations}\label{other_pipeline}
There are many possible configurations for an ASR pipeline. These can use
different types of voice-activity detection algorithms for
preprocessing (Figure~\ref{fig:pipeline}(b)) or any variety of feature
extraction algorithms (Figure~\ref{fig:pipeline}(c)), which include
DFTs, MFCCs, or Convolutional
Blocks. Similarly, an ASR system can use any number of model
types for inference, including DNN-HMM~\cite{kaldi_dnn},
DNN-RNN~\cite{ds2_pytorch,mozilla_ds} and
HMM-GMM~\cite{kaldi_hmm,sphinx}. ASRs with
different configurations are frequently introduced. 
Given the popularity of neural networks, we expect non-neural network VPS configurations to
eventually be phased out. As a result, the focus of our paper is
adversarial ML in the space of neural network based VPSes. This allows us to apply our findings to a larger population of VPSes.

\subsection{Speaker Identification (SI)}\label{si}
\subsubsection{Types}
SI systems can be broadly classified into two types: identification and verification. Identification systems determine the identity of the speaker of a given voice sample. In contrast, verification system ascertains whether the claimed identity of a speaker matches the given voice sample. For simplicity of exposition, we refer to both as SI.
\subsubsection{ASR vs SI}\label{asr_vs_si}
The modern ASR and SI pipelines are very similar to each other~\cite{abdullah2019kenensville}. Both of them use the overall structure illustrated in Figure~\ref{fig:ctc}. Both systems employ pre-processing, feature extraction, and inference. However, they differ at the last stage of the pipeline. While ASRs have a decoding stage (Figure~\ref{fig:ctc}(e)), SIs do not. Instead, an SI directly outputs the probability distribution across all the speakers effectively stopping at Figure~\ref{fig:ctc}(d).

\subsection{Adversarial Machine Learning in VPSes}
\label{ssec:aml}

The term adversarial machine learning refers to the study
of attacks and defenses for ML systems~\cite{Huang2011}. 
Attacks may target security
properties that relate to the integrity or confidentiality of the 
ML system~\cite{papernot2016towards}. 
The former encompasses poisoning attacks at training time~\cite{rubinstein2009antidote} and 
evasion attacks~\cite{biggio2013evasion} at test time. The latter may be concerned with the
confidentiality of data~\cite{ohrimenko2016oblivious} (often also delving into issues that relate to 
the privacy of data~\cite{dwork2006calibrating,song2019privacy,nasr2019comprehensive}) or the confidentiality of the model itself with attacks including
model extraction~\cite{tramer2016stealing,wang2018stealing}.  

We focus on
evasion attacks, which are often instantiated using maliciously
crafted inputs known as \textit{adversarial samples}~\cite{szegedy2013intriguing}. Most of the
literature on adversarial samples has been written in the context of
computer vision. In this realm, adversarial samples are produced
by introducing perturbations that do not affect semantics of the image
(as validated by a human observer) but that cause machine
learning models to misclassify the image. All existing attacks in 
the audio space are evasion attacks, which naturally
results in our systematization of knowledge addressing evasion
attacks. We do however discuss other types attacks in Section~\ref{discussion}.

\subsubsection{Motivating Example}\label{motivating_example}
Consider the example of an ASR. The goal of the adversary is to perturb an audio sample such that an ASR and a human transcribe the same audio differently. This perturbed audio sample, for example, can be music~\cite{yuan2018commandersong}, noise~\cite{biometrics}, silence~\cite{zhang2017dolphinattack} or even human speech~\cite{carlini2018audio}. This type of attack can trick humans into thinking that the perturbed sample is benign, but can force the voice-enabled home assistant to execute illicit bank transactions, unlock smart doors, etc.
Similarly, in the case of SI, the attacker might want to perturb an audio file so that the SI misidentifies the speaker.

The goal of the adversary is to find an algorithm capable of producing
perturbations imperceptible
to a human listener, yet still triggers the desired action from the 
voice assistant.
This would ensure that victims only realize they have been attacked
once the damage is done, if they realize at all.

\subsubsection{Crafting Adversarial Samples}\label{simple_attack}
Attack algorithms craft adversarial samples by performing gradient-based
optimization. Because it is possible to approximate
gradients using finite-difference methods sufficiently for the purpose
of finding adversarial samples~\cite{chen2017zoo}, we describe this
attack strategy assuming that the adversary has access to the model's gradients.

The adversary solves an optimization problem whose objective
is forcing the model to output a different label, subject to constraints that ensure the modified
input maintains the semantics that correspond to the original label.
\begin{equation}
\label{eq:advx}
\min_{\|\delta\|} l(f(x+\delta), y) + c \cdot \|\delta\|
\end{equation}
This formulation, due to Szegedy et al.~\cite{szegedy2013intriguing},
involves a loss function $l:x,y \mapsto l(f(x), y)$ measuring how far the model $f$'s
predictions are from label $y$ for input $x$.
The adversary who's optimizing for a targeted transcription needs to find
a targeted adversarial sample $x^*=x+\delta$ 
for which model $f$ erroneously outputs the label $y$, which is chosen by the adversary
and differs 
from the correct label assigned to input $x$. If instead the 
adversary is interested in crafting an untargeted adversarial sample,
they replace $l(f(x), y)$ by its negation and set $y$ to be the 
correct label of $x$. This encourages the procedure to find an input
$x^*$ that maximizes the model's error, regardless of the label
predicted by model $f$. The penalty $c \cdot \|\delta\|$ loosely
translates the requirement that an input should not be perturbed
excessively, otherwise it will no longer retain semantics that justify
it belonging in its original ground truth class.

\begin{figure}
  \includegraphics[width=\linewidth]{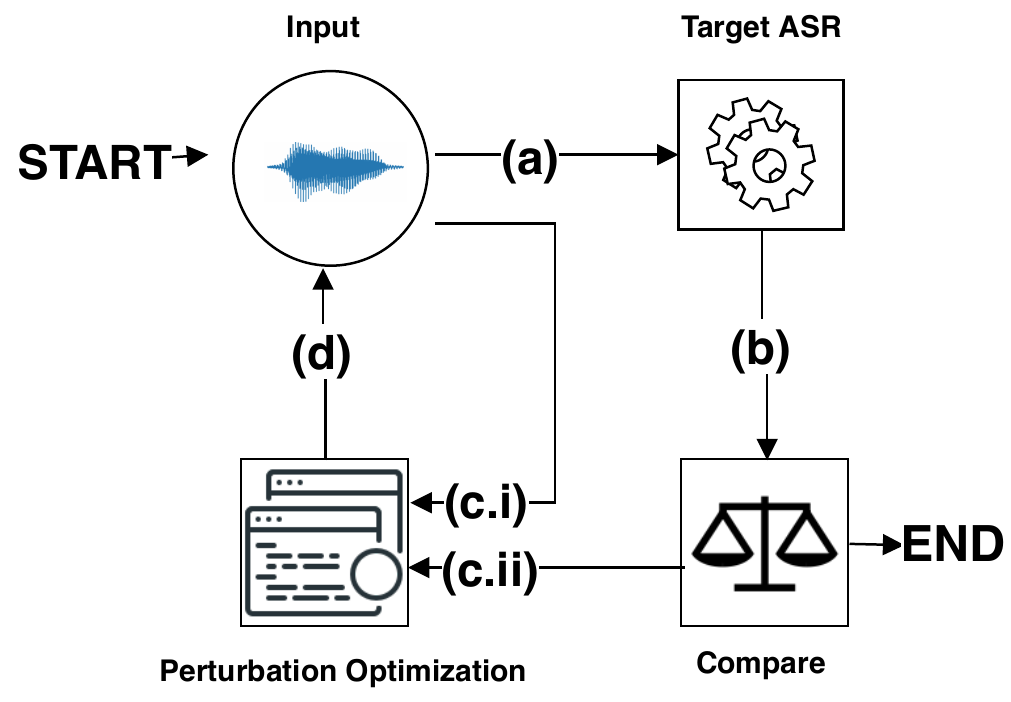}
  \caption{All iterative adversarial attacks follow the same general
algorithmic steps. (a) An input is passed to the model. (b) The Target
Model processes the input and outputs a label. (c) If the label does
not match the adversaries chosen label, the input (c.i) and the label
(c.ii) are passed to the perturbation optimization algorithm (d). The algorithm produces
a new adversarial input. These steps are repeated until the Target
Model outputs an adversary chosen label. } 
\label{fig:simple_attack}
\vspace{-0.8em}
\hrulefill
\vspace{-2.0em}
\end{figure}

For non-convex models such as neural networks, this optimization is approximated. Fortunately for the adversary, properties of 
neural networks that make them easy to train with algorithms like
stochastic gradient descent also result in neural networks being easy
to attack. In practice, this means that the adversary can solve
the optimization problem, formulated in the previously discussed equation, using
optimizers commonly employed to train neural networks, such
as gradient descent~\cite{papernot2016limitations} or some of its variants like
L-BFGS~\cite{szegedy2013intriguing} and Adam~\cite{carlini2017towards}. 
This often takes the form of a procedure iteratively modifying the input
until it is misclassified (Figure~\ref{fig:simple_attack}).

\subsubsection{Transferability}\label{transferability}\label{adv_sample_transfer}
When an adversarial sample that was crafted to be misclassified by Model A is
also misclassified by Model B, even though models A and B are
different, we say that the adversarial sample \textit{transfers} from 
model A to B~\cite{papernot2016transferability}. 
The two models A and B do not necessarily need to share the same architecture
and weights, nor do they need to be trained using the same technique
and training data. A hypothesis for transferability is that there is a
large volume of error space found in different models, which 
leaves a large volume when these error spaces 
intersect~\cite{tramer2017space}.
Transferability of adversarial samples is a
cornerstone of most attacks in the image domain, because the property
was
observed to hold in
even highly dissimilar
models~\cite{papernot2016transferability,demontis2019adversarial,jagielski2019high,wu2018understanding,liu2016delving}.
An adversary can perform an attack 
without having access to model B by first training a surrogate
model A. The surrogate is trained by querying the remote model B. Now the adversary can transfer adversarial samples from model A to model B. 
This reduces the knowledge required for attack success. 
\section{Attacks Against VPSes}
\label{signal_processing_attack}

One might assume it is possible to extend the existing 
literature of attacks from the image space to the audio domain.
This is not the case due to several key
differences. We detail these differences next. 
They motivate our systematization of knowledge effort
tailored to the audio domain in general and VPSes in particular.

\subsection{Pre-processing pipeline}
Speech recognition pipelines differ significantly from those employed
in image recognition because they pre-process data before it is analyzed by 
an ML model (Section~\ref{asr_pipe}). While image classifiers almost always operate directly
on raw pixels that encode the image, audio classifiers often rely on
feature extraction components that are not learned from the
training data. Instead, these features are extracted using signal processing algorithms
that are hard-coded by 
human experts~\cite{sigurdsson2006mel}. 

This difference enables a whole new class of audio adversarial samples that target
discrepancies between feature extraction techniques and the manner in which humans 
represent audio signals. These attacks use signal processing 
techniques to achieve the same evasion goals as the optimization-based
attacks presented in Section~\ref{ssec:aml}. Because these VPS attacks target
feature extraction, they are less model-dependent
and work as well as evasion attacks in the image domain even with limited access to 
the victim model~\cite{biometrics}.

\subsection{Sequential models}

Since audio is sequential in nature, ML models that solve
audio tasks are learned from architectures that are stateful in order to capture
contextual dependencies in an audio sample. 
Recall that neural networks for image recognition have
neurons arranged in layers where connections between neurons are
only made between layers. In contrast, \textit{recurrent} neural architectures support connections between neurons in the same layer in order to capture context. This enables the model
to propagate patterns identified in prior time steps
as it progresses temporally through the audio sample~(Section~\ref{infer}). 

Although recurrent neural architectures enable  accurate analysis of audio signals, 
the temporal dimension increases the difficulty
of performing a successful attack against VPSes. To craft 
a perturbation, the adversary now needs to consider all time
steps simultaneously. This is referred to as \textit{unrolling} in the ML literature. While unrolling enables one to backpropagate through time, it complicates optimization---with
failures such as exploding or vanishing gradients~\cite{pascanu2013difficulty}.
Because gradient-based algorithms rely on this optimization to converge,
finding adversarial samples is more challenging.

\subsection{Discrete domains}

Audio models had to be adapted
to the discrete nature of language. Whereas images define
a continuous input domain (because pixels are often represented
as three floating point values in ML systems),
language is represented as a sequence of tokens. Each
\textit{token} corresponds to a word included in the language's
dictionary or a character of the language's alphabet. Furthermore, some audio models are ``sequence-to-sequence'', they take in a sequence of words (e.g., as audio) and 
produce another sequence of words (e.g., as text). They must
handle the discrete nature of language both at their inputs
and outputs.

This introduces particularities when modeling
text or speech to adapt to discrete domains. For instance,
tokens are often projected in a continuous space of smaller dimensionality
than the size of the discrete dictionary with a technique called
\textit{word embeddings}~\cite{mikolov2013efficient}. This helps expose the relationship
between concepts expressed by different words.

For models defended with gradient masking~\cite{papernot17}
in the image domain, the presence of non-differentiable operations
like beam search constrains the adversary to come up with alternative
operations that are differentiable. Otherwise, they face
a large increase in the cost of finding adversarial samples~\cite{athalye2018obfuscated} or, in the worst case, have to resort to brute
force~\cite{athalye2018obfuscated}. This means that an attack might not always produce an adversarial sample that transcribes to the desired target text.

\subsection{Statistical models beyond ML}

Despite being replaced gradually with neural networks, some
components of voice-processing systems continue to involve
techniques like Hidden Markov Models~\cite{kaldi_hmm}. There is little work on attacking
these statistical models, and for this reason, it is not well
understood to what extent they are vulnerable to adversarial 
samples. More work needs to adapt gradient-based optimization attacks 
that rely on algorithms derived from the backpropagation algorithm to the expectation-maximization
algorithm used with Hidden Markov Models. This opens a new
dimension of attacks, because several alternatives to Hidden Markov Models
exist e.g., CTC~\cite{ds2_pytorch}.

\section{Attack Threat Model Taxonomy}\label{attack_threat_model}
We introduce the first 
threat model to characterize the unique contributions and open
problems of
evasion attacks in the audio domain. 
We hope this framework will help draw fair comparisons
between works published on VPSes in the future.

\subsection{Adversarial Goals}\label{perspective}


 We can group the attacker's goals into two categories of attacks: untargeted and targeted attacks.

\subsubsection{Untargeted}\label{untargeted_misclassification}
Here, the attacker wants the VPS to produce any output that is different from the original output.

	\paragraph{SI}
The attacker's goal is to force the SI to misidentify the speaker of an audio sample. If the model identifies the speaker of an audio sample as anyone other than the original, the attacker wins.
	\paragraph{ASR}
Similarly, the attacker aims to mislead the ASR into assigning any transcription to the input other
than the correct one. Consider the motivating example from
Section~\ref{motivating_example}, the ASR transcribes the original audio as \texttt{LOCK}.  
The adversary modifies the audio sample to produce an
adversarial sample. If the ASR transcribes the adversarial sample as anything other than \texttt{LOCK}, the attacker wins.

\subsubsection{Targeted}\label{targeted_misclassification}
A targeted attack is one where the attacker wants to get a specific response from the VPS.

	\paragraph{SI} In the case of an SI, the attacker goal is to force it to identify the attacker's chosen person as the speaker of the audio. For example, the attacker wants the SI to believe that the audio belongs to Bob (or that the audio sample contains silence), even though it does not.

	\paragraph{ASR} Similarly, in the case of an ASR, the attacker attempts to force the model to mistranscribe the input audio to a chosen
transcription. Consider the ASR from Section~\ref{motivating_example},
the attacker wants the ASR to assign the label~\texttt{OPEN} to the audio
sample. Compared to untargeted attacks, targeted ones can be harder to achieve
when sequence pairs have closer semantics than others. For example, it
may be easier to force the model to mistranscribe the audio for
the word~\texttt{LOCK} to~\texttt{CLOCK} than to~\texttt{OPEN}.

The only difference between targeted and untargeted attacks is the attacker's intentions towards the ASR/SI. If the intent is to force a specific output, then the attack is targeted. If however, the goal is to get any output, as long as it is not the correct one, then the attack is untargeted.

\subsection{Types of Adversarial Attacks}\label{adversarial_attacks}
Current attacks can be categorized as follows:

\subsubsection{Optimization Attacks}
\paragraph{Direct}\label{optimization_attacks}
Attacks use the information about the weights of the model to compute the gradients. These are then used to perturb the original input to move it in the decision space.

\paragraph{Indirect}\label{oracle_attacks}
These attacks generate adversarial samples by gradient estimation. This is done, indirectly, by repeatedly querying the target model~\cite{chen2017zoo}. Making enough queries to the model will reveal the underlying gradients and will help the attacker craft a working sample.

\subsubsection{Signal Processing Attacks}\label{signal_processing_atk}
These attacks use signal processing techniques to achieve the same goals as traditional optimization attacks. The signal processing attack techniques exploit discrepancies between the human ear and feature extraction algorithms. These attacks do not directly target the inference component of the pipeline, but nevertheless can still force the inference component to make incorrect predictions. Leveraging this divergent behavior enables signal processing attacks to be faster, more query efficient, and less model dependent compared to their optimization counterparts~\cite{biometrics}.

\subsubsection{Miscellaneous Attacks}\label{misc_attack}
These are the attacks that do not fall into any of the above two categories. These can exploit the VPS by targeting the limitations of the hardware (e.g., the microphone) or adding random noise. The range of miscellaneous attacks is very broad. For example, there is extensive work that has been done in the space of replay attacks~\cite{lau2018alexa,diao2014your,young2016badvoice}. Here an attacker captures the voice of a victim and attempts to exploit the VPS by replaying it the captured audio. Similarly, there are a number of attacks that exploit the inability of ASRs to distinguish between homophones: words that are spelled differently but sound the same (e.g., flour vs flower)~\cite{erichennenfentskill,zhang2019dangerous}. This happens when a single homophone is passed alone to the ASR. For example, just passing the word flower/flour, instead of as part of a sentence `which flowers grow in the summer'. Having a sentence provides context that allows the ASR to successfully differentiate the homophones. However, these attacks do not exploit any specific part of the VPS pipeline and are therefore outside the scope of this paper.

\subsection{Adversarial Knowledge}\label{attacker_knowledge}
A stronger attack is one that assumes the adversary has little
knowledge of the VPS components. A VPS is comprised of a few different components (Figure~\ref{fig:pipeline}). For the purposes of simplicity, we
will group these components into five categories.

\subsubsection{Component Categories}\label{component_categories}
VPS components include  task, preprocessing, feature extraction, inference,
and decoding. Information about a single component in one category, implies
knowledge of \textit{all} other components in the \textit{same} category.
\begin{compactitem}

    \item {\textit{Task:}} The problem the model is trained to solve (e.g.,
speech transcription) and data the model was trained on. For example, if an
adversary is attacking an English language transcription model, she is
aware that the system has been trained on an English language data-set. 
    
    \item {\textit{Preprocessing:}}
	A VPS first preprocesses the audio 
	(Figure~\ref{fig:pipeline}(b)). Knowledge of this step includes information about the
algorithm and parameters being used for down-sampling, noise reduction and
low-pass filtering.

    \item {\textit{Feature Extraction:}}
	As discussed in Section~\ref{asr_pipe}, a VPS pipeline is made up of
multiple signal processing steps for feature extraction. Knowledge of this component (Figure~\ref{fig:pipeline}(c)), allows the attacker to infer the feature extraction algorithm used by the VPS.

    \item {\textit{Inference:}}
	This step outputs a probability distribution over the labels
 (Figure~\ref{fig:pipeline}(d)). This category includes knowledge of
the weights, type, number of the layers and architecture of the model used for inference.
	
	 \item {\textit{Decoding:}}
	This step (Figure~\ref{fig:pipeline}(e))
converts the probability distribution into a human readable transcription. This
category includes information about the decoding algorithm (e.g., beam
search) and its parameters. 

\end{compactitem}

Having defined the components, we now categorize the different knowledge types
(Table~\ref{tab:knowledge}):

\paragraph{White-Box}\label{white_box}
The attacker has perfect knowledge of all the above categories, even
though an attack may exploit only a specific component. A white-box attack is
the best case scenario for the attacker. An example this access type is an
open source model (e.g., DeepSpeech-2~\cite{ds2_pytorch} or
Kaldi~\cite{kaldi_hmm}~\cite{kaldi_dnn}).

\paragraph{Grey-Box}\label{grey_box}
The attacker has knowledge of only a subset of the categories. She might have
complete knowledge of some components, and limited or no knowledge of others.
An example is Azure Speaker Recognition
model~\cite{azure_verify}~\cite{azure_attest}. The information about this
system's filter extraction and task category is publicly
available~\cite{sadjadi2013msr}. However, there is no information about the
preprocessing, inference and decoding components are unknown to the public.


\paragraph{Black-Box}\label{black_box}
The attacker has knowledge of only the task category. For example, the attacker
only knows that the target system is an English language transcription model.
This information is generally available for any public VPS. An example of this
is Google Speech API~\cite{google_normal}, Amazon Alexa~\cite{amazon} and
Siri~\cite{siri}. In contrast, the preprocessing,
feature extraction, inference and decoding components are all unknown. 


\paragraph{No-Box}\label{no_box}
This is an extreme version of the black-box access type. The attacker has no
knowledge of any of the categories. Consider the example of the attacker
attempting to subvert a telephony surveillance system. Infrastructure for such
a surveillance system uses VPSes to efficiently convert millions of hours of
captured call audio into searchable text. The attacker does not have access to
this infrastructure. She only knows
that her phone calls will be captured and transcribed. Of all the knowledge
types, this is the most restrictive. A no-box setting is
the worst-case scenario for the attacker. 

\begin{table}

\begin{tabular}{c|c|c|c|c|c}
\textbf{Knowledge} & \textbf{Task} & \shortstack{\textbf{Pre-}\\ \textbf{processing}} & \shortstack{\textbf{Feature}\\ \textbf{Extract}} & \textbf{Inference} & \textbf{Decoding} \\
\hline
\textbf{White-Box} & \cmark & \cmark & \cmark & \cmark & \cmark \\
\textbf{Grey-Box} & \textbf{?} & \textbf{?} & \textbf{?} & \textbf{?} & \textbf{?} \\
\textbf{Black-Box} & \cmark & \xmark & \xmark & \xmark & \xmark \\
\textbf{No-Box} & \xmark & \xmark & \xmark & \xmark & \xmark \\

\end{tabular}
\caption{Different categories of knowledge available to the attacker.
``\xmark'' information unavailable to the attacker.
``\cmark'' information available to the attacker. \textbf{``?''} information may or may not be available to the attacker.}
\label{tab:knowledge}
\vspace{-0.8em}
\hrulefill
\vspace{-2.0em}
\end{table}

\subsection{Adversarial Capabilities}\label{capabilities}

\subsubsection{Constrains on the Input Manipulations}\label{input_maniputlation}

\paragraph{Input and Output Granularity}\label{input_granularity}
Attacks against VPSes are of three granularities: phoneme\footnote{A phoneme is a single distinct unit of sound in a language.}, word and sentence
level. These measure two aspects about the attack: the window of input that the attacker will need to perturb and total change in output transcription. For example, consider an attacker who can mistranscribe an entire sentence by only changing a single phoneme. Here, the input granularity is phoneme and the output granularity is sentence. Similarly, if the attacker needs to perturb the entire sentence to change the transcription of a single word, then input granularity is sentence and the output granularity is word.

\paragraph{Types of Adversarial Audio}\label{audio_type}
Depending on the attack type and scenario, adversaries can produce different
types of audio samples. These audio samples can be categorized into the
following broad classes:

\begin{compactitem}
	\item \textit{Inaudible: }\label{inaudible}
	As discussed in Section~\ref{hearing_range}, the human auditory system can
	only perceive frequencies that range from 20Hz to 20kHz. In contrast, by using
	microphones to capture audio, VPSes can record frequencies beyond 20kHz. To
	exploit this discrepancy, attackers can encode an audio command in the
	ultrasound frequency range (20kHz to 10Mhz). The encoded command is detected
	and recorded by the microphone but is inaudible to the human listener. The
	recorded command is then passed onto the VPS which executes it, considering it
	human speech. These inaudible attacks have been able to exploit modern VPS
	devices such as Alexa, Siri, and Google Home~\cite{zhang2017dolphinattack}.
	These attacks can not be filtered out in software, as they exploit the hardware
	component (microphone) of the pipeline. The audio sample is aliased down to less
	than 20kHz when recorded by the microphone. However, these attacks might still
	leave artifacts in the signal that can potentially reveal the attack. While attacks
	in this space focus on the ultrasound frequencies, the frequency range below 20Hz is
	also inaudible and could be a vector in this class of attack.
	\item \textit{Noise: }\label{noise}
	This category of attacks produces audio samples that sound like noise to
	humans but are considered legitimate audio commands by the VPS. It is difficult
	for the human auditory system to interpret audio that is non-continuous,
	jittery and lossy. Attacks can mangle audio samples such that they will not be
	intelligible by humans and will (hopefully) be ignored as mere noise. However,
	the same mangled audio sample is processed by the VPS as legitimate speech.
	This can allow attackers to trick a VPS into unauthorized action. 
	\item \textit{Clean: }\label{clean}
	The last category of attacks perturb audio such that it sounds clean to humans
	even though there is a hidden command embedded inside it. These attacks embed
	commands as low-intensity perturbations, into an audio sample such as music,
	which is not noticeable by humans. However, the VPS detects and executes these
	embedded commands. 
\end{compactitem}

\subsubsection{Access to the Model}

\paragraph{Queries}\label{query_count}
A query consists of sending the model an input and receiving the corresponding
output. As described in Section~\ref{simple_attack}, the attacker makes
multiple queries to a target model to produce a single adversarial sample. A
threat model must consider the maximum number of times an attacker can query the
victim system. Too many queries will alert the defender of an attack. Additionally, most proprietary systems require users to
pay for every query. Too many queries can quickly compound the
monetary cost of an attack. 

\paragraph{Output}\label{target_model_output}\label{probability_distribution}\label{label}
A model can produce one of two types of outputs. This can either be a single
\textit{label} (i.e., transcription), or a \textit{probability distribution} over all the labels. 

A probability distribution over all the labels is the individual probabilities
of the input belonging to each of the labels. For example, the modle might output the following
distribution for a recording of ``lock'': \texttt{LOCK:} 90\%, \texttt{LOOK:} 7\% and \texttt{LOKK:} 3\%. This
means that the model is 90\% , 7\%, and 3\% certain that the input is ``lock'',
``look'', and ``lokk'' respectively. 


The model is still calculating the probability distribution when it outputs a
single label. However, the model will only return the label with the highest
probability. It is entirely possible for the adversary to have access to some combination of both label and  probability distribution. For example, she may have the final label and
the next top K labels.

\subsubsection{Attack Medium}\label{attack_medium}
\begin{figure}
  \centering
  \includegraphics[width=\linewidth]{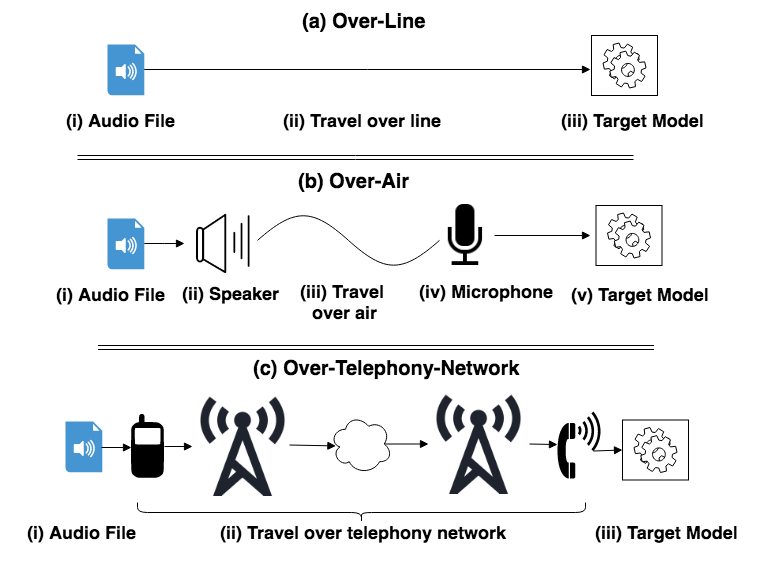}
  \caption{Depending on the threat model, an attack audio might be passed over various mediums.
  (a) Over-Line Attack: Audio file is passed directly to the target model as a \texttt{.wav} file. (b) Over-Telephony-Network Attack: The audio transmitted over the telephony network to the target model. (c) Over-Air Attack: The audio file is played using a speaker. The audio travels over the air and is recorded by a microphone. The recorded signal is then passed onto the target model.}
\label{fig:over_mediums}
\vspace{-0.8em}
\hrulefill
\vspace{-2.0em}
\end{figure}

Depending on the scenario, the adversarial sample can be passed to the system over different mediums,
each of which can introduce new challenges (e.g, noise). In the case of adversarial images, the
perturbed input can be passed directly as a .jpg file. However, in a more realistic
scenario, the attacker will have to print the perturbed image, and hope when
the target takes a picture of the image that the perturbations are captured by
the camera. In the case of VPSes, the mediums can be broadly grouped into four
types: Over-Line (Figure~\ref{fig:over_mediums}(a)), Over-Air (Figure~\ref{fig:over_mediums}(b)),
Over-Telephony-Network (Figure~\ref{fig:over_mediums}(c)), and
Over-Others. In this section, we discuss each of the mediums in detail.


\paragraph{Over-Line}\label{over_line}
The input is passed to the model directly as a Waveform Audio file or
\texttt{.wav} (Figure~\ref{fig:over_mediums}(a)). Compared to all the other mediums,
attacks Over-Line are the easiest to execute as this medium ensures ``lossless"
transmission. 

\paragraph{Over-Air}\label{over_Air}
Over-Air (Figure~\ref{fig:over_mediums}(b)), involves playing the audio using a
speaker. As an example, consider an attacker who wants to exploit an Amazon
Alexa. The attacker plays the adversarial audio over the speaker, which travels
through the air. The audio is then recorded and interpreted by Alexa. As
mentioned earlier, adversarial perturbations are sensitive and can be
lost during transmission. In the Over-Air scenario, the loss can occur due to
interference, background noise or imperfect acoustic equipment~\cite{blue2018hello}. Therefore,
attacks that produce audio that can survive the highly lossy mediums of
air are much stronger than those that cannot.

\paragraph{Over-Telephony-Network}\label{over_telephony_network}
An attack Over-Telephony-Network (Figure~\ref{fig:over_mediums}(c)), involves
playing audio samples over the telephony network and exploiting any VPS
transcribing the phone call. The telephony network is a lossy medium due to
static interference, codec compression, packet loss, and
jitter~\cite{bordonaro2005method,bordonaro2005method,keshav1997engineering,rix2001perceptual,pesola1993electromagnetic,bolot1996control,reaves2016authloop}.
Adversarial audio is sensitive to lossy mediums and is at risk of losing
perturbations during transmission. Therefore, attacks that produce audio
samples that can survive the telephony network are also much stronger than those
that cannot.

\paragraph{Over-Others}\label{over_others}
This category includes mediums that do not fall into the above three
categories. One example of this is MPEG-1 Audio Layer III or \texttt{mp3}
compression. The \texttt{.wav} audio samples are often compressed using
\texttt{mp3} compression before being transmitted. This compression technique
is lossy and is bound to result in some of the adversarial perturbations being
discarded. Attacks that produce perturbations that only survive
\texttt{mp3} compression are stronger than Over-Line, but are weaker than
Over-Telephony-Network and Over-Air.

\subsubsection{Distance}
The further an attack audio sample travels over a lossy medium, the greater the degradation. As a result, the attack audio sample is less likely to exploit the target VPS. One metric to measure the strength of an attack is the distance it can travel without losing its exploitative nature.

\subsubsection{Acoustic Environment}
This includes the different acoustic environments that the attack audio sample was tested in (e.g., noisy environments). This is because the various mediums introduce unique sources of distortion. For Over-Air attacks, this involves testing the attack audio sample with varying levels of noise. For Over-Telephony-Network attacks, this involves testing the attack over real telephony networks.

\subsubsection{Acoustic Equipment}
Over-Air and Over-Telephony-Network attacks involve passing the audio samples over a range of acoustic equipment. These include codecs, microphones, speakers, etc., which are subject to hardware imperfections. Certain frequencies might be attenuated and intensified, based on the particular equipment's impulse response. However, if an attack audio sample does not work against a range of equipment, then it is too sensitive. Any attack that makes assumptions tied to a particular set of acoustic equipment is considered weaker than one that does not.

\section{Existing Attack Classification}\label{overview}
We use the threat model taxonomy described in the previous section to classify the existing attacks and
identify the areas that need improvement. Table~\ref{tab:current_overview_1} and
Table~\ref{tab:current_overview_2} show the progress of current attacks against VPSes.

\begin{table*}
\begin{tabular}{cccccccccc}
\textbf{Attack Name} & \textbf{Audio Type} & \textbf{Goal} & \shortstack{\textbf{Input} \\
\textbf{Granularity}} & \shortstack{\textbf{Output} \\
\textbf{Granularity}} & \textbf{Knowledge} & \textbf{Queries} & \textbf{Output} & \textbf{Medium} & \textbf{Time} \\
Taori et al.~\cite{taori2018targeted} & Clean & Targeted & S & S & Gray & \textbf{?} & Distribution & L & \textbf{?} \\
Carlini et al.~\cite{carlini2018audio} & Clean & Targeted & S & S & White & 1000 & Distribution & L & \textbf{?} \\
Houdini~\cite{cisse2017houdini} & Clean & Targeted & S & S & White & \textbf{?} & Distribution & L & \textbf{?} \\
Kreuk et al.~\cite{kreuk2018fooling} & Clean & Targeted & S & S & White & \textbf{?} & Distribution & L & \textbf{?} \\
Qin et al.~\cite{qin2019imperceptible} & Clean & Targeted & S & S & White & \textbf{?} & Distribution & L & \textbf{?} \\
Schonherr et al.~\cite{schonherr2018adversarial} & Clean & Targeted & S & S & White & 500 & Distribution & L & \textbf{?} \\
Abdoli et al.~\cite{abdoli2019universal} & Clean & Targeted & W & W & White & \textbf{?} & Distribution & L & \textbf{?} \\
Commander Song~\cite{yuan2018commandersong} & Clean & Targeted & S & S & White & \textbf{?} & Distribution & L,A & \textbf{?} \\
Yakura et al.~\cite{yakura2018robust} & Clean & Targeted & S & S & White & \textbf{?} & Distribution & L,A & 18 hours \\
Devil's Whisper~\cite{chen2020devil} & Clean & Targeted & S & S & Black & 1500 & Label & L,A & 4.6 hours \\
M. Azalnot et al.~\cite{alzantot2018did} & Clean & Targeted & W & W & Black & \textbf{?} & Label & L & \textbf{?} \\
Kenansville Attack~\cite{abdullah2019kenensville} & Clean & Untargeted & P,W,S & P,W,S & No & 15 & Label & L,T & ~seconds \\
Dolphin Attack~\cite{zhang2017dolphinattack} & Inaudible & Targeted & S & S & Black & \textbf{?} & Label & A & \textbf{?} \\
Light Commands~\cite{sugawara2020light} & Inaudible & Targeted & S & S & Black & \textbf{?} & Label & A & \textbf{?} \\
Abdullah et al.~\cite{biometrics} & Noisy & Targeted & S & S & Black & 10 & Label & L,A & ~seconds \\
Cocaine Noodles~\cite{vaidya2015cocaine} & Noisy & Targeted & S & S & Black & \textbf{?} & Label & L,A & \textbf{?} \\
HVC (2)~\cite{carlini2016hidden} & Noisy & Targeted & S & S & Black & \textbf{?} & Label & L,A & 32 hours \\
HVC (1)~\cite{carlini2016hidden} & Noisy & Targeted & S & S & White & \textbf{?} & Distribution & L & \textbf{?} \\
\end{tabular}

\caption{The table\protect\footnotemark~ provides an overview of the current progress of the adversarial attacks against ASR and SI systems. ``\textbf{?}'': Authors provide no information in paper. ``\cmark'': Will work. ``P,W,S'' = Phoneme, Word, Sentence. ``L,A,T'' = Over-Line, Over-Air, Over-Telephony-Network.}

\label{tab:current_overview_1}
\vspace{-1em}
\end{table*}
\footnotetext{The most up-to-date version of this table can be found at: https://sites.google.com/view/adv-asr-sok/}

\begin{table*}

\begin{tabular}{cccccccccc}
\textbf{Attack Name} & \shortstack{\textbf{Attack} \\
\textbf{Type}} & \shortstack{\textbf{VPS} \\
\textbf{Attacked}} & \shortstack{\textbf{VPS} \\
\textbf{Internals}} & \textbf{ASR} & \textbf{SI} & \shortstack{\textbf{Distance} \\
\textbf{(Approximate)}} & \shortstack{\textbf{Acoustic} \\
\textbf{Equipment}} & \shortstack{\textbf{Acoustic} \\
\textbf{Environment}} & \textbf{Transferability} \\
Carlini et al.~\cite{carlini2018audio} & Direct & 1 & RNN & \cmark & \xmark\fminus & \xmark\fminus & \xmark\fminus & \xmark\fminus & \xmark\fminus \\
HVC (1)~\cite{carlini2016hidden} & Direct & 1 & RNN & \cmark & \xmark\fminus & \xmark\fminus & \xmark\fminus & \xmark\fminus & \xmark\fminus \\
Houdini~\cite{cisse2017houdini} & Direct & 1 & RNN & \cmark & \cmark\fplus & \xmark\fminus & \xmark\fminus & \xmark\fminus & \xmark\fminus \\
Schonherr et al.~\cite{schonherr2018adversarial} & Direct & 1 & HMM & \cmark & \xmark & \xmark\fminus & \xmark\fminus & \xmark\fminus & \xmark\fminus \\
Kreuk et al.~\cite{kreuk2018fooling} & Direct & 1 & RNN & \xmark\fminus & \cmark\fplus & \xmark\fminus & \xmark\fminus & \xmark\fminus & \xmark\fminus \\
Qin et al.~\cite{qin2019imperceptible} & Direct & 1 & RNN & \cmark & \xmark\fminus & \xmark\fminus & \xmark\fminus & \xmark\fminus & \xmark\fminus \\
Yakura et al.~\cite{yakura2018robust} & Direct & 1 & RNN & \cmark & \cmark\fminus & 0.5 meters & \cmark & \xmark\fminus & \xmark\fminus \\
Commander Song~\cite{yuan2018commandersong} & Direct & 2 & HMM & \cmark & \xmark\fminus & 1.5 meters & \cmark & \textbf{?} & \cmark \\
Devil's Whisper~\cite{chen2020devil} & Direct & 3 & \textbf{?} & \cmark & \xmark\fminus & 5-200 cm & \cmark & \cmark & \cmark \\
Abdoli et al.~\cite{abdoli2019universal} & Direct & 5 & CNN & \cmark & \xmark\textbf{?} & \xmark\fminus & \xmark\fminus & \xmark\fminus & \xmark\fminus \\
M. Azalnot et al.~\cite{alzantot2018did} & Indirect & 1 & CNN & \cmark & \xmark & \xmark & \xmark & \xmark & \xmark \\
Taori et al.~\cite{taori2018targeted} & Indirect & 1 & RNN & \cmark & \xmark\textbf{?} & \xmark\textbf{?} & \xmark\textbf{?} & \xmark\textbf{?} & \xmark\textbf{?} \\
HVC (2)~\cite{carlini2016hidden} & Miscellaneous & 1 & \textbf{?} & \cmark & \xmark\fminus & 0.5 meters & \xmark\fminus & \xmark\fminus & \xmark\fminus \\
Cocaine Noodles~\cite{vaidya2015cocaine} & Miscellaneous & 1 & \textbf{?} & \cmark & \xmark\fminus & 30 cm & \xmark\fminus & \xmark\fminus & \xmark\fminus \\
Dolphin Attack~\cite{zhang2017dolphinattack} & Miscellaneous & 9 & RNN,\textbf{?} & \cmark & \xmark & 150 cm & \cmark & \cmark & \cmark \\
Light Commands~\cite{sugawara2020light} & Miscellaneous & 4 & \textbf{?} & \cmark & \cmark & 110 m & NA & NA & \cmark \\
Kenansville Attack~\cite{abdullah2019kenensville} & Signal Processing & 6 & RNN,\textbf{?} & \cmark & \cmark & N/A & \cmark & \cmark & \cmark \\
Abdullah et al.~\cite{biometrics} & Signal Processing & 12 & RNN,HMM,\textbf{?} & \cmark & \cmark & 1 ft & \cmark & \cmark & \cmark \\
\end{tabular}

\caption{The table shows the current progress of the adversarial attacks against ASR and SI systems.``\textbf{?}'': Authors provide no information in paper.``\cmark'': Will work. We sent each of the authors of the above papers emails regarding their papers and have included the responses with  ``\xmark'' in the table. ``\xmark\textbf{?}'': Authors did not test it and are not sure if it will work.  ``\xmark\fplus'': Authors did not test it and believe it will work. ``\xmark\fminus'': Authors did not test it and believe it will not work.``\xmark'': Authors did not respond to correspondence but we believe it will not work.}
\label{tab:current_overview_2}
\vspace{-0.8em}
\hrulefill
\vspace{-2.0em}
\end{table*}

\subsubsection{Targeted White-box Attacks}
Most targeted attacks in Table~\ref{tab:current_overview_1} require complete white-box
knowledge of the
target~\cite{yuan2018commandersong,carlini2018audio,cisse2017houdini,schonherr2018adversarial,kreuk2018fooling,qin2019imperceptible,abdoli2019universal,yakura2018robust}.
All of these are based on the optimization strategy and can embed hidden commands
inside audio samples such as music or noise. These attacks can generate
adversarial samples that are either clean or noisy.

One of two conditions must be satisfied for these attacks to be applicable in the real-world. Either the attacker has perfect knowledge of the target or the adversarial samples are transferable. However, both these assumptions are impractical. Regarding the former, it is unlikely for an attacker to have perfect knowledge of the target VPS, primarily because these systems are proprietary. For example, the underlying construction and functionality of ASRs such as Alexa and Siri are a closely guarded trade secret. Similarly, the latter does not hold either, as we will discuss in Section~\ref{transferability_expts}. As a result, these white-box attacks are constrained against real-world systems like Alexa or Siri.

White-box attacks have made significant contributions in the space of adversarial ML. However, the authors generally do not provide metrics to help identify the advantages of their attacks over existing ones. These include metrics such as max number of queries, time to produce an adversarial audio sample etc. Generally, the attack methods make similar assumptions (e.g., input/output granularity, goals, knowledge, medium etc.), but exploit different ASR architectures. For example, Carlini et al.~\cite{carlini2018audio} exploit DeepSpeech (CNN-RNN), while Schonherr et al.~\cite{schonherr2018adversarial} and Commander Song~\cite{yuan2018commandersong} both target Kaldi (DNN-HMM).

\subsubsection{Clean Attacks and Mediums}\label{over-air-fail}
Clean audio attacks produce audio samples that sound clear to humans but are mistranscribed by the ASR (Section~\ref{capabilities}). Although these attacks have been demonstrated Over-Line, they have only seen limited success over other mediums (Table~\ref{tab:current_overview_2}). There are only a few attacks that can work Over-Air~\cite{yuan2018commandersong,qin2019imperceptible,yakura2018robust}. These are constrained as they require white-box knowledge of the target, physical access to the victim's acoustic equipment, can take hours to generate, have limited physical range and are sensitive to background noise. Similarly, there is only a single attack~\cite{abdullah2019kenensville} that can function Over-Telephony-Network. However, it can not produce a targeted transcription.

Success Over-Line does not translate into success over other mediums. There are a plethora of factors that contribute to this limitation (Section~\ref{attack_medium}). These interfere with the adversary generated perturbations, effectively blunting the attack. Though these factors have a negligible impact on the ASR understanding of a benign audio sample, these can frustrate the efforts of the attacker. To enable clean attacks Over-Air and Over-Telephony-Network, researchers can benefit from characterizing and overcoming the various sources of interference.

\subsubsection{Signal Processing Attacks vs Rest}

In comparison to the other attack types, signal processing attacks have shown greater promise, specifically in black-box settings. These attacks have achieved the same goals as traditional optimization attacks, specifically for noise~\cite{biometrics} and untargeted attacks~\cite{abdullah2019kenensville}. However, unlike their counterparts~\cite{carlini2016hidden}, signal processing attacks are more efficient as they require less than 15 queries and a few seconds to generate an attack sample. These attacks assume black-box access, which has allowed them to be successfully tested against more target VPSes than existing works. However, clean targeted attacks based on signal processing have not yet been demonstrated.

\subsubsection{Attacks against SIs}

Only a tiny subset of the existing work has focused on attacking SI systems~\cite{kreuk2018fooling, biometrics, abdullah2019kenensville} (Table~\ref{tab:current_overview_2}). This is a result of the similarity of the underlying architectures of the SIs and ASRs. Generally, an attack against ASR can be used wholesale to exploit an SI. This is true for both signal processing and optimization attacks. This is because, as discussed in Section~\ref{asr_vs_si}, pipelines for systems often constitute the same stages. If an attack exploits the stage in the pipeline shared by both ASRs and SIs, it will succeed against both. This is corroborated by the examples of the same attacks being successfully used, as is, against both systems. Signal processing attack papers~\cite{biometrics, abdullah2019kenensville} have already demonstrated this ability to exploit both ASRs and SIs.

\subsubsection{Indirect Optimization Attacks}

Indirect optimization attacks attempt to generate adversarial samples by repeatedly querying the model~\cite{alzantot2018did,taori2018targeted} in a black-box setting (Table~\ref{tab:current_overview_2}). These attacks have been demonstrated only at the word~\cite{alzantot2018did} granularity. One might incorrectly assume that indirect attacks can be used against real-world systems (e.g., Alexa) as these do not require perfect knowledge of the target, but these attacks have limited for a number of reasons. 
First, these attacks require hundreds of thousands of queries to generate a single attack audio sample. This makes attack execution difficult as most ASR's charge users a fee for each query~\cite{amazon}. Second, the audio samples generated using these attacks can \textit{only} be used over-line and have not been demonstrated over other mediums. This is primarily due to the sensitivity of the samples to loss/distortion (Section~\ref{attack_medium}). Third, these attacks have not been demonstrated at the sentence level in a black-box setting. The only attack~\cite{taori2018targeted} that comes close still requires information about the distributions, and is therefore grey box. Overcoming these limitations is a direction for future research.



\subsubsection{Optimization Attacks Do Not Guarantee Success}

Optimization attacks explore the decision spaces using algorithms like
gradient descent. One well-known drawback of such algorithms is that they can get stuck
in local minima. This means that it might not be possible to successfully
perturb every benign audio sample to evade the target VPS. In contrast,
signal processing attacks have been free of this specific limitation. These
attacks have been able to guarantee attack success for any benign audio sample.

\subsubsection{Model Agnostic Attacks}

 Only a small subset of existing attacks methods are black-box~\cite{alzantot2018did,vaidya2015cocaine,zhang2017dolphinattack,biometrics,abdullah2019kenensville,taori2018targeted} (Table~\ref{tab:current_overview_2}) and therefore model agnostic, while the remaining attacks are model dependent as each exploits a unique component of the target VPS. For example, Carlini et al. ~\cite{carlini2018audio} exploit the CTC component of the DeepSpeech ASR~\cite{ds1_tf}, which does not exist in the Kaldi ASR~\cite{kaldi_dnn}. It is important for future researchers to develop model agnostic attacks, since newer ASR systems with distinct architectures and improved accuracies are being introduced every day. In this highly variable and continuously changing space, any attack that only works against a single VPS type can quickly become obsolete.

\section{Defense and Detection Taxonomy}\label{def_threat_model}
We present a taxonomy to help categorize the space of defenses and detection methods.

\begin{table*}
\begingroup
\setlength{\tabcolsep}{5pt} 
\renewcommand{\arraystretch}{1} 
\begin{tabular}{cccccccccccc}
\textbf{Name} & \textbf{Stochastic Model} & \shortstack{\textbf{Additional} \\
\textbf{Hardware}} & \shortstack{\textbf{Attacker} \\
\textbf{Type}} & \textbf{Distance} & \shortstack{\textbf{Attack} \\
\textbf{Type}} & \shortstack{\textbf{Audio} \\
\textbf{Type}} & \textbf{Medium} & \textbf{ASR} & \textbf{SI} & \multicolumn{2}{c}{ \textbf{Adversarial Cost} } \\
& & & & & & & & & & \textbf{Resources} & \textbf{Distortion} \\
Wang et al. \cite{wang2019secure} & \cmark & \cmark & \textbf{?} & \textbf{?} & Miscellaneous & N/A & A & \textbf{?} & \textbf{?} & \textbf{?} & \textbf{?} \\
VoicePop \cite{wang2019voicepop} & \cmark & \xmark & \textbf{?} & 6 cm & Miscellaneous & N/A & A & \textbf{?} & \textbf{?} & \textbf{?} & \textbf{?} \\
Blue et al. \cite{blue2018hello} & \xmark & \xmark & ADPT, N-ADPT & 5 m & Miscellaneous & N/A & A & \textbf{?} & \textbf{?} & \textbf{?} & \textbf{?} \\
Wang et al. \cite{wang2019defeating} & \cmark & \cmark & \textbf{?} & \textbf{?} & Miscellaneous & Noisy & A & \textbf{?} & \textbf{?} & \textbf{?} & \textbf{?} \\
Yang et al. \cite{yang2018characterizing} & \xmark & \xmark & N-ADPT & N/A & Direct & Clean & L & \cmark & \xmark & \textbf{?} & \textbf{?} \\
\end{tabular}

\caption{The table\protect\footnotemark~ provides an overview of the current defenses for ASR and SI systems. ``\textbf{?}'': Authors provide no information in paper. ``\xmark'': Does not work or has not been demonstrated. ``\cmark'': Will work. ``P,W,S'' = Phoneme, Word, Sentence. ``L,A,T'' = Over-Line, Over-Air, Over-Telephony-Network. ``ADPT, N-ADPT'' = effective against Adaptive Attacker, effective against Non-Adaptive Attacker. }\label{tab:def_overview}

\endgroup
\vspace{-0.8em}
\hrulefill
\vspace{-2.0em}
\end{table*}

\subsection{Attacker Type:}
There are two types of attackers:
\subsubsection{Non-adaptive} 
This attacker does not have any knowledge of the target's defense strategy and parameters. The minimal criteria a viable defense strategy needs to meet is to be able prevent non-adaptive attackers.

\subsubsection{Adaptive} 
The attacker has full knowledge of the the type and parameters of the defense strategy being employed. Using this knowledge, the attacker can modify their attack strategy in hopes of overcoming the defender. A strong defense can prevent successful exploitation even in the presence of an adaptive attacker.

\subsection{Adversarial Cost:}
\subsubsection{Resources}
A defense might force an attacker to expend greater resources to produce adversarial audio samples. For example, an attacker would be forced to make additional queries to the model. A defense is strong if it can increase the resources required for a viable attack by a significant margin.

\subsubsection{Distortion}
The goal of an attacker is to control the degree of audible distortion introduced due to perturbations. For example, in order to circumvent VPSes used in telephony surveillance system, the adversary has the dual aims of evading the VPS and ensuring low audible distortion. Large distortions might make the audio message difficult to understand. A defense strategy might aim to force the attacker to increase the audible distortion needed to successfully create an attack sample. Adding too much distortion will make the attack audio sample difficult to understand, preventing the second aim. A strong defense will force the attacker to add significant amounts of additional distortion to the sample, such that it becomes a determent to their original aims. However, measure audible distortion is difficult, due to lack of any psychoacoustic metric, as discussed later in Section~\ref{Intelligibility_Metrics}.

\subsection{Stochastic Modeling:}
Defenses against these attacks can often use stochastic (or ML) models as part of the pipeline. Such techniques gather different features that might help differentiate an adversarial sample from a benign one. These features are then used to train a ML model to do the classification. A strong defense does not rely on ML models as these models are themselves vulnerable to exploitation.

\subsection{Additional Hardware:}
Some defense techniques might require sensors in addition to the microphone. This is problematic as it increases the cost of manufacturing the home assistants. Due to already thin profit margins~\cite{asr_cost}, additional sensor cost will decrease the likelihood of manufacturers incorporating the defense. A strong defense does not incur any additional deployment costs.

\subsection{Distance:}
When defending against Over-Air attacks, some techniques are only effective if the source of the adversarial audio is within a certain distance from the target. This is because the defense techniques might use certain identifying features, which might otherwise be lost during transmission. The longer the distance, the stronger the defense.

It is worth discussing the ideal distance. Over-Air attacks require an audio sample to be played over a speaker that is present within the same room as the target VPS. Within the US, rooms are on average 11ft by 16ft~\cite{room_size}. This means in the worst case, the attacker will need to play an audio file from the farther part of the room. If the target VPS is in one corner of the room and the attacker's speaker in the other corner, this distance will come down to 20ft (Pythagorean Theorem).  

\subsection{Attack Type:}
This is the specific type of attack the strategy is supposed to protect against (Section~\ref{adversarial_attacks}). These include signal processing attacks, direct optimization and indirect optimization attacks. A strong defense is universal i.e., stops any type of attack.

\subsection{Audio Type:}
This is the specific type of audio that strategy is designed to defeat (Section~\ref{audio_type}). These include clean, inaudible, and noisy audio. A strong defense is universal i.e., can stop an adversarial audio of any type.
\footnotetext{The most up-to-date version of this table can be found at: https://sites.google.com/view/adv-asr-sok/}

\subsection{Medium:}
The medium the strategy is designed to defend (Section~\ref{attack_medium}). A strong defense can stop an attack over any medium (Over-Air, Over-Telephony, and Over-Line).

\section {Defenses and Detection Classification}\label{defense}

There has been little published work in the space of defenses and detection mechanisms for adversarial audio. In this section, we discuss how the most popular defense strategy from the adversarial image space, adversarial training, is not effective in the audio domain. We also analyze the published mechanism for detecting adversarial audio and propose a direction for future research.

\subsection{Adversarial training}\label{def:adversrail_training}
This defense involves training the model on samples perturbed using adversarial algorithms. This strategy has shown promise in the adversarial image space~\cite{madry2017towards}. Intuitively, this improves the decision boundary, by either making it more robust to attacks or by making adversarial samples harder to craft by obfuscating the gradients. To exploit the adversarially trained model, the attacker will either need to run the attack algorithm for more iterations or introduce greater distortion to the adversarial input. We argue that this is not be an effective method for defending VPSes. 


First, adversarial training can decrease the accuracy of the model. This phenomenon is known as label leaking~\cite{kurakin2016adversarial}. Here, the adversarially trained model shows improved robustness to adversarial samples, but at the cost of decreased accuracy on legitimate samples. Given that VPSes (e.g. Siri, Amazon Alexa) are user-facing, reducing their accuracy will degrade the user experience and consequently, the vendor's profits. Second, adversarial training might not work against the signal processing attacks discussed in 
Section~\ref{signal_processing_atk}. These generate perturbations that are filtered out 
during feature extraction~\cite{biometrics,zhang2017dolphinattack}. Thus, both adversarial and legitimate samples will collide to similar feature vectors. As a result, adversarial training might decrease model accuracy with no improvement to robustness~\cite{abdullah2019kenensville}. This is different from \emph{label leaking} as the model accuracy for both benign and adversarial samples will suffer.

\subsection{Liveness Detection}\label{def:detection}\label{def:liveness_detection}
Considering the limitations of adversarial training, it is important to discuss detection strategies that have shown some success. Detection mechanisms are designed to help identify whether an audio sample is benign or malicious. Detection often suffers from the same limitations as adversarial training: for instance, both fare poorly against adaptive adversaries.
Liveness detection is an area of research that aims to identify whether the source of speech was a real human or a mechanical speaker. It can help prevent replay audio attacks i.e., an attacker plays an audio of someone ordering something back to their home assistant. This is a popular research area with yearly competitions~\cite{todisco2019asvspoof}. Considering this is still an open research area, we only cover a handful of important works, shown in Table~\ref{tab:def_overview}.

Liveness detection can help detect Over-Air attacks as they require playing the audio over mechanical speakers. However, works in this area limited to a number of reasons. First, these works~\cite{wang2019secure,wang2019voicepop,wang2019defeating} employ ML models as part of the pipeline. This adds another layer of vulnerability as attackers have the ability to exploit ML models using adversarial algorithms. This means an adaptive attacker, with knowledge of the detection method, can overcome it. Second, these works either require the distance between the source and target to be very small ~\cite{wang2019voicepop} or the authors fail to disclose this number at all Table~\ref{tab:def_overview}~\cite{wang2019secure,wang2019defeating}. An attacker can merely execute the attack from a distance further than the one ideal of the detection method to work. One mechanism that has shown promise is Blue et al. ~\cite{blue2018hello}. This defense is effective at a much larger distance and is not defeated by an adaptive attacker.



\subsection{Future Direction}

It is worth discussing future directions that may not guarantee success. An example of this is redesigning the feature extraction and preprocessing stages. Signal processing attacks are thus far unique to ASR systems. These attacks exploit limitations and vulnerabilities in the preprocessing and feature extraction phases of the ASR pipeline. Even though the existence of this class of adversarial examples is clear, these are still not easy to resolve. This is because the techniques that are used during feature extraction (e.g., the DFT) have taken decades to develop~\cite{rabiner1978digital}. These techniques are well understood by the research community and have significantly improved the accuracy of ASR systems. Developing new techniques that not only resolve current security flaws but also maintain high ASR accuracy is difficult and does not guarantee robustness.

\section{Discussion}\label{discussion}

In this section, we highlight key findings, discuss their implications and make recommendations for future research.

\subsubsection{Lack of Transferability for Optimization Attacks}
Transferability has been shown for signal processing attacks. However until now, this question was largely unanswered for the case of optimization attacks. In this paper, we empirically demonstrate that transferability of optimization attacks is unlikely in the audio domain. In fact, this is true even if both surrogate and remote target models share the same architecture, hyperparameters, random seed and training data. This is a result of training on GPUs, which introduces non-determinism that can lead to the VPSes learning different decision boundaries~\cite{jooybar2013gpudet}. Therefore, for adversarial audio samples to transfer, the adversary must not only have complete information of the model's training parameters, but also be able to predict the GPU-induced randomness during the training process.
We believe that this latter requirement is unrealistic for practical adversaries.

The only optimization attack that has demonstrated transferability, although in a limited sense, is Commander Song~\cite{yuan2018commandersong}. The authors were able to transfer samples generated for the Kaldi ASR~\cite{kaldi_dnn} to iFlytek~\cite{kaldi_iflytek}, but failed when transferring the samples to DeepSpeech~\cite{mozilla_ds_0.4.1}. A likely hypothesis is that iFlytek is using a fine-tuned version of the same pre-trained Kaldi ASR, as has been the case in the past~\cite{kaldi_iflytek}. Thus a caveat to the lack of transferability is that adversarial samples may be transferred between a fine-tuned model and its pre-trained counterpart. This could be explained by the fact that decision boundaries do not change significantly during the fine-tuning process. Additionally, an increasing number of vendors are transitioning to NN based systems, away from the HMMs that the Kaldi ASR employs internally. Therefore, the question of lack of transferability for optimization attacks should be considered more seriously with regards to NN based ASRs.



\subsubsection{Defenses for VPSes}
A number of defenses have been proposed for the computer vision domain~\cite{cretu2008casting,steinhardt2017certified,meng2017magnet,jordaney2017transcend,biggio2015one,bendale2016towards,xu2017feature,tramer2017ensemble,lecuyer2019certified}. These have been~\textit{primarily} designed to defend against adversaries who might exploit transferability. However, the transferability is difficult in the audio domain, and attackers rarely have white-box access to target VPSes (e.g., Amazon Alexa). This minimizes the threat of existing white-box attacks against real world systems. Consequently, researchers should focus on building defenses against attacks that have been demonstrated in black-box settings, such as signal processing attacks~\cite{biometrics,abdullah2019kenensville}.

\subsubsection{VPSes Pipeline}
The modern VPS pipeline is completely different from that of image models, due to presence of additional components. Each of these components increases the attack surface, introducing a unique set of vulnerabilities that an attacker may be able to exploit. The full scope of vulnerabilities has yet to be uncovered, with some attacks (e.g., clean, targeted attacks) not having been demonstrated. Therefore, future research should focus on identifying and exploiting novel weaknesses within the pipeline. Similarly, \textit{if the entire pipeline can be attacked, then the entire pipeline needs to be 
defended}. Thus, we recommend that future research efforts focus on building robust defenses for each individual component of the pipeline.

\subsubsection{Lack of Poisoning and Privacy Attacks}
This paper focuses on evasion attacks to the detriment of other adversarial machine learning attacks such as poisoning attacks and privacy attacks~\cite{papernot2016towards}. This is because, to the best of our knowledge, no poisoning attacks or privacy attacks have been proposed for speech. And existing attacks which may apply have not been evaluated. This is an interesting direction for future research to explore. Poisoning attacks generate audio samples that, when added to the training data, make the model misbehave in an attacker-controlled way. For example, poisoning can be used prevent the model from correctly transcribing certain types of inputs. In contrast, privacy attacks attempt to uncover information about the model's training data. For example, an attacker may want to determine if the voice of a certain individual was used for training a speaker identification system.

\subsubsection{Detection Mechanisms}

While observing Table~\ref{tab:def_overview}, an astute reader might have realized that no mechanism yet exists to defend against telephony-attacks. while liveness detection and temporal based mechanisms have demonstrated some success in addressing Over-Air and Over-Line attacks. These methods are not perfect, but still constitute a positive step towards addressing these attacks. In stark contrast, there is no work that addresses Over-Telephony attacks. Given that these attacks can be most reliably executed against real world surveillance systems, it would be ideal to focus research efforts in this space.

\subsubsection{Lack of Audio Intelligibility Metrics}\label{Intelligibility_Metrics}
A number of methods have been used to measure intelligibility of audio. However, these methods have limitations. Researchers have used metrics from the computer vision domain such as the L2-norm~\cite{carlini2018audio}. This is not an adequate metric to measure audio intelligibility as the human ear does not exhibit linear behavior (Section~\ref{acoustics}). Thus, audio samples that are jarring to the human ear, can still have small L2-norm~\cite{biometrics}. In addition, prior work often includes users studies that measure the quality of attack audio samples~\cite{cisse2017houdini,carlini2016hidden,abdullah2019kenensville,qin2019imperceptible}. Unfortunately, these studies do not  consider the full range of variables that impact human intelligibility. These include age~\cite{select-1, age}, first language~\cite{fam}, audio equipment, hearing range~\cite{TED} and environmental noise. Future works should consider these variables to improve generalizability of their findings. Finally, some researchers use audio quality metrics designed for the telephony netowrks. These are metrics are designed to measure audio quality of telephony lines~\cite{rix2001perceptual,itu2018863} and effectively measure quantities like jitter, packet loss, and white-noise; which are facets of audio that existing attacks do not target.



Psychoacoustics is a promising direction for designing suitable audio intelligibility metrics. Recall from 
Section~\ref{perception_mechanism} that human perception of speech is affected by a combination of mechanisms. Some of these mechanisms are quantitative and can be used to construct hearing 
models~\cite{schroeder1975models}. An example is audio masking~\cite{lin2015principles}. Attack perturbations are, in effect, frequencies that have been introduced to the benign audio samples. Some of these frequencies will mask other frequencies. This masking effect can be measured using metrics such as tone-to-noise ratio and prominence ratio. These metrics, in combination with other metrics from an hearing model, can help measure the quality of an adversarial speech sample. 

\section{Conclusion}\label{conc}
Modern VPSes use neural networks to convert audio samples into text (ASRs) or identify the speaker (SIs). However, neural networks have been shown to be vulnerable to adversarial machine learning attacks.  These can force VPSes to act maliciously. In this paper, we present a threat model framework to evaluate existing works in the adversarial space against VPSes. We identify the unique contributions, open problems and future research directions.

The space of attacks against VPSes is different and more complex than that of their image counterparts. This is because VPSes are comprised of additional phases including processing, feature extraction, and decoding algorithms. This means attacks against image models cannot be easily extended to VPSes. This has lead to the development of attacks designed for VPSes. However, most of these attacks can not be used against real-world systems. This is primarily due to lack of success in black-box settings, failure Over-Air, and limited transferability. There has also been limited development of adequate defenses for VPSes. While there are a plethora of defenses and detection mechanisms in the image domain, only one exists for VPSes, which is limited to optimization attacks.

%
\bibliographystyle{IEEEtran}
\bibliography{bib/refs,bib/defs,bib/misc}

\begin{thebibliography}{100}
\providecommand{\url}[1]{#1}
\csname url@samestyle\endcsname
\providecommand{\newblock}{\relax}
\providecommand{\bibinfo}[2]{#2}
\providecommand{\BIBentrySTDinterwordspacing}{\spaceskip=0pt\relax}
\providecommand{\BIBentryALTinterwordstretchfactor}{4}
\providecommand{\BIBentryALTinterwordspacing}{\spaceskip=\fontdimen2\font plus
\BIBentryALTinterwordstretchfactor\fontdimen3\font minus
  \fontdimen4\font\relax}
\providecommand{\BIBforeignlanguage}[2]{{%
\expandafter\ifx\csname l@#1\endcsname\relax
\typeout{** WARNING: IEEEtran.bst: No hyphenation pattern has been}%
\typeout{** loaded for the language `#1'. Using the pattern for}%
\typeout{** the default language instead.}%
\else
\language=\csname l@#1\endcsname
\fi
#2}}
\providecommand{\BIBdecl}{\relax}
\BIBdecl

\bibitem{CM17}
C.~Martin, ``{72\% Want Voice Control In Smart-Home Products},'' Accessed in
  2019, available at
  \url{https://www.mediapost.com/publications/article/292253/2-want-voice-control-in-smart-home-products.html?edition=99353}.

\bibitem{HS18}
H.~Stephenson, ``{UX design trends 2018: from voice interfaces to a need to not
  trick people},'' Accessed in 2019, available at
  \url{https://www.digitalartsonline.co.uk/featurs/interactive-design/ux-design-trends-2018-from-voice-interfaces-need-not-trick-people/}.

\bibitem{article_acc_models}
``{Who's Smartest: Alexa, Siri, and or Google Now?}'' Last accessed in 2019,
  available at \url{https://bit.ly/2ScTpX7}.

\bibitem{azure_attest}
``{Azure Speaker Identification API},'' Last accessed in 2019, available at
  \url{https://azure.microsoft.com/en-us/services/cognitive-servic/speaker-recognition/}.

\bibitem{mozilla_ds}
``{Mozilla Project DeepSpeech},'' Last accessed in 2019, available at
  \url{https://azure.microsoft.com/en-us/services/cognitive-servic/speaker-recognition/}.

\bibitem{carlini2018audio}
N.~Carlini and D.~Wagner, ``Audio adversarial examples: Targeted attacks on
  speech-to-text,'' in \emph{2018 IEEE Security and Privacy Workshops
  (SPW)}.\hskip 1em plus 0.5em minus 0.4em\relax IEEE, 2018, pp. 1--7.

\bibitem{biometrics}
H.~Abdullah, W.~Garcia, C.~Peeters, P.~Traynor, K.~Butler, and J.~Wilson,
  ``{Practical Hidden Voice Attacks against Speech and Speaker Recognition
  Systems},'' \emph{Proceedings of the 2019 Network and Distributed System
  Security Symposium (NDSS)}, 2019.

\bibitem{lin2015principles}
Y.~Lin and W.~H. Abdulla, ``Principles of {P}sychoacoustics,'' in \emph{Audio
  Watermark}.\hskip 1em plus 0.5em minus 0.4em\relax Springer, 2015, pp.
  15--49.

\bibitem{google}
``{Cloud Speech-to-Text},'' Last accessed in 2019, available at
  \url{https://cloud.google.com/speech-to-text/}.

\bibitem{amazon}
``{Amazon Lex},'' Last accessed in 2019, available at
  \url{https://aws.amazon.com/?nc2=h_lg}.

\bibitem{siri}
``{Apple's Siri},'' Last accessed in 2019, available at
  \url{https://www.apple.com/siri/}.

\bibitem{NSA_1}
``{The Computers are Listening, Part 2},'' Accessed in 2018, available at
  \url{https://theintercept.com/2015/06/nsa-transcription-american-phone-calls/}.

\bibitem{NSA_2}
``{The Computers are Listening, Part 1},'' Accessed in 2018, available at
  \url{https://theintercept.com/2015/05/nsa-speech-recognition-snowden-searchable-text/}.

\bibitem{NSA_3}
\BIBentryALTinterwordspacing
``{For 10 Years, NSA Software Has Made Phone Calls Searchable by Keyword},''
  Accessed in 2018. [Online]. Available:
  \url{"https://www.truthdig.com/articl/for-10-years-nsa-software-has-made-phone-calls-searchable-by-keyword/"}
\BIBentrySTDinterwordspacing

\bibitem{NSA_4}
``{The Computers are Listening, Part 1},'' Accessed in 2018, available at
  \url{https://theintercept.com/2015/06/nsa-transcription-american-phone-calls/}.

\bibitem{NSA_5}
``{Report: NSA Can Record, Store Phone Conversations of Whole Countries},''
  Accessed in 2018, available at
  \url{https://www.npr.org/sections/thetwo-way/2014/03/18/2911652/report-nsa-can-record-store-phone-conversations-of-whole-countries}.

\bibitem{otter}
``{Otter.ai},'' Last accessed in 2019, available at
  \url{https://otter.ai/login}.

\bibitem{sohn1999statistical}
J.~Sohn, N.~S. Kim, and W.~Sung, ``A statistical model-based voice activity
  detection,'' \emph{IEEE signal processing letters}, vol.~6, no.~1, pp. 1--3,
  1999.

\bibitem{sigurdsson2006mel}
S.~Sigurdsson, K.~B. Petersen, and T.~Lehn-Schi{\o}ler, ``{Mel Frequency
  Cepstral Coefficients: An Evaluation of Robustness of MP3 Encoded Music.}''
  in \emph{ISMIR}, 2006, pp. 286--289.

\bibitem{rabiner1978digital}
L.~R. Rabiner and R.~W. Schafer, \emph{Digital processing of speech
  signals}.\hskip 1em plus 0.5em minus 0.4em\relax Prentice Hall, 1978.

\bibitem{ahmed1974discrete}
N.~Ahmed, T.~Natarajan, and K.~R. Rao, ``Discrete cosine transform,''
  \emph{IEEE transactions on Computers}, vol. 100, no.~1, pp. 90--93, 1974.

\bibitem{gunning2017explainable}
D.~Gunning, ``Explainable artificial intelligence (xai),'' \emph{Defense
  Advanced Research Projects Agency (DARPA), nd Web}, 2017.

\bibitem{hannun2017sequence}
A.~Hannun, ``Sequence modeling with ctc,'' \emph{Distill}, vol.~2, no.~11,
  p.~e8, 2017.

\bibitem{koehn2004pharaoh}
P.~Koehn, ``Pharaoh: a beam search decoder for phrase-based statistical machine
  translation models,'' in \emph{Conference of the Association for Machine
  Translation in the Americas}.\hskip 1em plus 0.5em minus 0.4em\relax
  Springer, 2004, pp. 115--124.

\bibitem{kaldi_dnn}
``{Kaldi ASpIRE Chain Model},'' Last accessed in 2019, available at
  \url{http://kaldi-asr.org/models.html}.

\bibitem{ds2_pytorch}
S.~Naren, ``Speech {R}ecognition using {DeepSpeech-2},'' Last accessed in 2019,
  available at \url{https://github.com/SeanNaren/deepspeech.pytorch}.

\bibitem{kaldi_hmm}
D.~Povey, A.~Ghoshal, G.~Boulianne, L.~Burget, O.~Glembek, N.~Goel,
  M.~Hannemann, P.~Motlicek, Y.~Qian, P.~Schwarz, J.~Silovsky, G.~Stemmer, and
  K.~Vesely, ``{The Kaldi Speech Recognition Toolkit},'' in \emph{IEEE 2011
  Workshop on Automatic Speech Recognition and Understanding}.\hskip 1em plus
  0.5em minus 0.4em\relax IEEE Signal Processing Society, 2011, iEEE Catalog
  No.: CFP11SRW-USB.

\bibitem{sphinx}
P.~Lamere, P.~Kwok, W.~Walker, E.~Gouvea, R.~Singh, B.~Raj, and P.~Wolf,
  ``{Design of the CMU Sphinx-4 decoder},'' in \emph{Eighth European Conference
  on Speech Communication and Technology}, 2003.

\bibitem{abdullah2019kenensville}
H.~Abdullah, M.~S. Rahman, W.~Garcia, L.~Blue, K.~Warren, A.~S. Yadav,
  T.~Shrimpton, and P.~Traynor, ``{Hear ``No Evil'', See ``Kenansville'':
  Efficient and Transferable Black-Box Attacks on Automatic Speech Recognition
  Systems},'' \emph{In Submission}, 2019.

\bibitem{Huang2011}
\BIBentryALTinterwordspacing
L.~Huang, A.~D. Joseph, B.~Nelson, B.~I. Rubinstein, and J.~D. Tygar,
  ``{Adversarial Machine Learning},'' in \emph{Proceedings of the 4th ACM
  Workshop on Security and Artificial Intelligence}, ser. AISec '11.\hskip 1em
  plus 0.5em minus 0.4em\relax New York, NY, USA: ACM, 2011, pp. 43--58.
  [Online]. Available: \url{http://doi.acm.org/10.1145/2046684.2046692}
\BIBentrySTDinterwordspacing

\bibitem{papernot2016towards}
N.~Papernot, P.~McDaniel, A.~Sinha, and M.~Wellman, ``Towards the science of
  security and privacy in machine learning,'' \emph{arXiv preprint
  arXiv:1611.03814}, 2016.

\bibitem{rubinstein2009antidote}
B.~I. Rubinstein, B.~Nelson, L.~Huang, A.~D. Joseph, S.-h. Lau, S.~Rao,
  N.~Taft, and J.~D. Tygar, ``{ANTIDOTE: Understanding and Defending against
  Poisoning of Anomaly Detectors},'' in \emph{Proceedings of the 9th ACM
  SIGCOMM conference on Internet measurement}.\hskip 1em plus 0.5em minus
  0.4em\relax ACM, 2009, pp. 1--14.

\bibitem{biggio2013evasion}
B.~Biggio, I.~Corona, D.~Maiorca, B.~Nelson, N.~{\v{S}}rndi{\'c}, P.~Laskov,
  G.~Giacinto, and F.~Roli, ``Evasion attacks against machine learning at test
  time,'' in \emph{Joint European conference on machine learning and knowledge
  discovery in databases}.\hskip 1em plus 0.5em minus 0.4em\relax Springer,
  2013, pp. 387--402.

\bibitem{ohrimenko2016oblivious}
O.~Ohrimenko, F.~Schuster, C.~Fournet, A.~Mehta, S.~Nowozin, K.~Vaswani, and
  M.~Costa, ``Oblivious {Multi-Party} {Machine} {Learning} on {Trusted}
  {Processors},'' in \emph{25th $\{$USENIX$\}$ Security Symposium
  ($\{$USENIX$\}$ Security 16)}, 2016, pp. 619--636.

\bibitem{dwork2006calibrating}
C.~Dwork, F.~McSherry, K.~Nissim, and A.~Smith, ``Calibrating {Noise} to
  {Sensitivity} in {Private} {Data} {Analysis},'' in \emph{Theory of
  cryptography conference}.\hskip 1em plus 0.5em minus 0.4em\relax Springer,
  2006, pp. 265--284.

\bibitem{song2019privacy}
L.~Song, R.~Shokri, and P.~Mittal, ``Privacy {Risks} of {Securing} {Machine}
  {Learning} {Models} against {Adversarial} {Examples},'' \emph{arXiv preprint
  arXiv:1905.10291}, 2019.

\bibitem{nasr2019comprehensive}
M.~Nasr, R.~Shokri, and A.~Houmansadr, ``Comprehensive {Privacy} {Analysis} of
  {Deep} {Learning}: Passive and {Active} {White-box} {Inference} {Attacks}
  against {Centralized} and {Federated} {Learning},'' in \emph{2019 IEEE
  Symposium on Security and Privacy (SP)}.\hskip 1em plus 0.5em minus
  0.4em\relax IEEE, 2019, pp. 739--753.

\bibitem{tramer2016stealing}
F.~Tram{\`e}r, F.~Zhang, A.~Juels, M.~K. Reiter, and T.~Ristenpart, ``Stealing
  {Machine} {Learning} {Models} via {Prediction} {APIs},'' in \emph{25th
  $\{$USENIX$\}$ Security Symposium ($\{$USENIX$\}$ Security 16)}, 2016, pp.
  601--618.

\bibitem{wang2018stealing}
B.~Wang and N.~Z. Gong, ``Stealing {Hyperparameters} in {Machine} {Learning},''
  in \emph{2018 IEEE Symposium on Security and Privacy (SP)}.\hskip 1em plus
  0.5em minus 0.4em\relax IEEE, 2018, pp. 36--52.

\bibitem{szegedy2013intriguing}
C.~Szegedy, W.~Zaremba, I.~Sutskever, J.~Bruna, D.~Erhan, I.~Goodfellow, and
  R.~Fergus, ``Intriguing properties of neural networks,'' \emph{arXiv preprint
  arXiv:1312.6199}, 2013.

\bibitem{yuan2018commandersong}
X.~Yuan, Y.~Chen, Y.~Zhao, Y.~Long, X.~Liu, K.~Chen, S.~Zhang, H.~Huang,
  X.~Wang, and C.~A. Gunter, ``{CommanderSong: A Systematic Approach for
  Practical Adversarial Voice Recognition},'' in \emph{{Proceedings of the
  USENIX Security Symposium}}, 2018.

\bibitem{zhang2017dolphinattack}
G.~Zhang, C.~Yan, X.~Ji, T.~Zhang, T.~Zhang, and W.~Xu, ``{DolphinAttack:
  Inaudible voice commands},'' in \emph{Proceedings of the 2017 ACM SIGSAC
  Conference on Computer and Communications Security}.\hskip 1em plus 0.5em
  minus 0.4em\relax ACM, 2017, pp. 103--117.

\bibitem{chen2017zoo}
P.-Y. Chen, H.~Zhang, Y.~Sharma, J.~Yi, and C.-J. Hsieh, ``{Zoo: Zeroth order
  optimization based black-box attacks to deep neural networks without training
  substitute models},'' in \emph{Proceedings of the 10th ACM Workshop on
  Artificial Intelligence and Security}.\hskip 1em plus 0.5em minus 0.4em\relax
  ACM, 2017, pp. 15--26.

\bibitem{papernot2016limitations}
N.~Papernot, P.~McDaniel, S.~Jha, M.~Fredrikson, Z.~B. Celik, and A.~Swami,
  ``The limitations of deep learning in adversarial settings,'' in
  \emph{Security and Privacy (EuroS\&P), 2016 IEEE European Symposium
  on}.\hskip 1em plus 0.5em minus 0.4em\relax IEEE, 2016, pp. 372--387.

\bibitem{carlini2017towards}
N.~Carlini and D.~Wagner, ``Towards evaluating the robustness of neural
  networks,'' in \emph{Security and Privacy (SP), 2017 IEEE Symposium
  on}.\hskip 1em plus 0.5em minus 0.4em\relax IEEE, 2017, pp. 39--57.

\bibitem{papernot2016transferability}
N.~Papernot, P.~McDaniel, and I.~Goodfellow, ``Transferability in {Machine}
  {Learning}: from {Phenomena} to {Black-Box} {Attacks} using {Adversarial}
  {Samples},'' \emph{arXiv preprint arXiv:1605.07277}, 2016.

\bibitem{tramer2017space}
F.~Tram{\`e}r, N.~Papernot, I.~Goodfellow, D.~Boneh, and P.~McDaniel, ``The
  {S}pace of {T}ransferable {A}dversarial {E}xamples,'' \emph{arXiv preprint
  arXiv:1704.03453}, 2017.

\bibitem{demontis2019adversarial}
A.~Demontis, M.~Melis, M.~Pintor, M.~Jagielski, B.~Biggio, A.~Oprea,
  C.~Nita-Rotaru, and F.~Roli, ``{W}hy {D}o {A}dversarial {A}ttacks {T}ransfer?
  {E}xplaining {T}ransferability of {E}vasion and {P}oisoning {A}ttacks,'' in
  \emph{28th $\{$USENIX$\}$ Security Symposium ($\{$USENIX$\}$ Security 19)},
  2019, pp. 321--338.

\bibitem{jagielski2019high}
M.~Jagielski, N.~Carlini, D.~Berthelot, A.~Kurakin, and N.~Papernot,
  ``High-{Fidelity} {Extraction} of {Neural} {Network} {Models},'' \emph{arXiv
  preprint arXiv:1909.01838}, 2019.

\bibitem{wu2018understanding}
L.~Wu, Z.~Zhu, C.~Tai \emph{et~al.}, ``Understanding and {Enhancing} the
  {Transferability} of {Adversarial} {Examples},'' \emph{arXiv preprint
  arXiv:1802.09707}, 2018.

\bibitem{liu2016delving}
Y.~Liu, X.~Chen, C.~Liu, and D.~Song, ``Delving into {Transferable}
  {Adversarial} {Examples} and {Black-box} {Attacks},'' \emph{arXiv preprint
  arXiv:1611.02770}, 2016.

\bibitem{pascanu2013difficulty}
R.~Pascanu, T.~Mikolov, and Y.~Bengio, ``On the difficulty of training
  {Recurrent} {Neural} {Networks},'' in \emph{International conference on
  machine learning}, 2013, pp. 1310--1318.

\bibitem{mikolov2013efficient}
T.~Mikolov, K.~Chen, G.~Corrado, and J.~Dean, ``Efficient {Estimation} of
  {Word} {Rrepresentations} in {Vector} {Space},'' \emph{arXiv preprint
  arXiv:1301.3781}, 2013.

\bibitem{papernot17}
N.~Papernot, P.~McDaniel, I.~Goodfellow, S.~Jha, Z.~B. Celik, and A.~Swami,
  ``{Practical Black-box Attacks Against Machine Learning},'' in
  \emph{Proceedings of the 2017 ACM on Asia Conference on Computer and
  Communications Security}.\hskip 1em plus 0.5em minus 0.4em\relax ACM, 2017,
  pp. 506--519.

\bibitem{athalye2018obfuscated}
A.~Athalye, N.~Carlini, and D.~Wagner, ``Obfuscated {G}radients {G}ive a
  {F}alse {S}ense of {S}ecurity: Circumventing {D}efenses to {A}dversarial
  {E}xamples,'' \emph{arXiv preprint arXiv:1802.00420}, 2018.

\bibitem{lau2018alexa}
J.~Lau, B.~Zimmerman, and F.~Schaub, ``Alexa, are you listening? privacy
  perceptions, concerns and privacy-seeking behaviors with smart speakers,''
  \emph{Proceedings of the ACM on Human-Computer Interaction}, vol.~2, no.
  CSCW, pp. 1--31, 2018.

\bibitem{diao2014your}
W.~Diao, X.~Liu, Z.~Zhou, and K.~Zhang, ``Your voice assistant is mine: How to
  abuse speakers to steal information and control your phone,'' in
  \emph{Proceedings of the 4th ACM Workshop on Security and Privacy in
  Smartphones \& Mobile Devices}, 2014, pp. 63--74.

\bibitem{young2016badvoice}
P.~J. Young, J.~H. Jin, S.~Woo, and D.~H. Lee, ``Badvoice: Soundless
  voice-control replay attack on modern smartphones,'' in \emph{2016 Eighth
  International Conference on Ubiquitous and Future Networks (ICUFN)}.\hskip
  1em plus 0.5em minus 0.4em\relax IEEE, 2016, pp. 882--887.

\bibitem{erichennenfentskill}
D.~Kumar, R.~Paccagnella, P.~Murley, E.~Hennenfent, J.~Mason, A.~Bates, and
  M.~Bailey, ``{Skill Squatting Attacks on Amazon Alexa},'' in \emph{27th
  {USENIX} Security Symposium ({USENIX} Security 18)}.\hskip 1em plus 0.5em
  minus 0.4em\relax {USENIX} Association, 2018.

\bibitem{zhang2019dangerous}
N.~Zhang, X.~Mi, X.~Feng, X.~Wang, Y.~Tian, and F.~Qian, ``Dangerous skills:
  Understanding and mitigating security risks of voice-controlled third-party
  functions on virtual personal assistant systems,'' in \emph{Dangerous Skills:
  Understanding and Mitigating Security Risks of Voice-Controlled Third-Party
  Functions on Virtual Personal Assistant Systems}.\hskip 1em plus 0.5em minus
  0.4em\relax IEEE, 2019, p.~0.

\bibitem{azure_verify}
``{Azure Speaker Verification API},'' Last accessed in 2019, available at
  \url{https://azure.microsoft.com/en-us/services/cognitive-servic/speaker-recognition/}.

\bibitem{sadjadi2013msr}
S.~O. Sadjadi, M.~Slaney, and L.~Heck, ``{MSR} identity toolbox v1. 0: A
  {MATLAB} toolbox for speaker-recognition research,'' \emph{Speech and
  Language Processing Technical Committee Newsletter}, vol.~1, no.~4, pp.
  1--32, 2013.

\bibitem{google_normal}
``{Google Cloud Speech-to-Text API},'' Last accessed in 2019, available at
  \url{https://cloud.google.com/speech-to-text/}.

\bibitem{blue2018hello}
L.~Blue, L.~Vargas, and P.~Traynor, ``Hello, is it me you're looking for?
  differentiating between human and electronic speakers for voice interface
  security,'' in \emph{Proceedings of the 11th ACM Conference on Security \&
  Privacy in Wireless and Mobile Networks}, 2018, pp. 123--133.

\bibitem{bordonaro2005method}
F.~G. Bordonaro, K.~Zhang, and S.~R. Raparla, ``Method and apparatus for
  measuring network data packet delay, jitter and loss,'' ~15 2005, uS Patent
  6,868,094.

\bibitem{keshav1997engineering}
S.~Keshav and S.~Kesahv, \emph{An engineering approach to computer networking:
  ATM networks, the Internet, and the telephone network}.\hskip 1em plus 0.5em
  minus 0.4em\relax Addison-Wesley Reading, 1997, vol.~1.

\bibitem{rix2001perceptual}
A.~W. Rix, J.~G. Beerends, M.~P. Hollier, and A.~P. Hekstra, ``Perceptual
  evaluation of speech quality (pesq)-a new method for speech quality
  assessment of telephone networks and codecs,'' in \emph{2001 IEEE
  International Conference on Acoustics, Speech, and Signal Processing.
  Proceedings (Cat. No. 01CH37221)}, vol.~2.\hskip 1em plus 0.5em minus
  0.4em\relax IEEE, 2001, pp. 749--752.

\bibitem{pesola1993electromagnetic}
M.~Pesola, T.~Saarnimo, V.-M. Valimaa, and A.~Leman, ``Electromagnetic
  interference shielding construction in a radio telephone,'' Dec.~14 1993, uS
  Patent 5,271,056.

\bibitem{bolot1996control}
J.-C. Bolot and A.~Vega-Garcia, ``Control mechanisms for packet audio in the
  internet,'' in \emph{Proceedings of IEEE INFOCOM'96. Conference on Computer
  Communications}, vol.~1.\hskip 1em plus 0.5em minus 0.4em\relax IEEE, 1996,
  pp. 232--239.

\bibitem{reaves2016authloop}
B.~Reaves, L.~Blue, and P.~Traynor, ``Authloop: End-to-end cryptographic
  authentication for telephony over voice channels,'' in \emph{25th USENIX
  Security Symposium USENIX Security 16}, 2016, pp. 963--978.

\bibitem{taori2018targeted}
R.~Taori, A.~Kamsetty, B.~Chu, and N.~Vemuri, ``{Targeted Adversarial Examples
  for Black Box Audio Systems},'' \emph{arXiv preprint arXiv:1805.07820}, 2018.

\bibitem{cisse2017houdini}
M.~Ciss{\'{e}}, Y.~Adi, N.~Neverova, and J.~Keshet, ``{Houdini: Fooling Deep
  Structured Visual and Speech Recognition Models with Adversarial Examples},''
  in \emph{Advances in Neural Information Processing Systems 30: Annual
  Conference on Neural Information Processing Systems 2017, 4-9 December 2017,
  Long Beach, CA, {USA}}, 2017, pp. 6980--6990.

\bibitem{kreuk2018fooling}
F.~Kreuk, Y.~Adi, M.~Cisse, and J.~Keshet, ``{Fooling End-to-end Speaker
  Verification by Adversarial Examples},'' \emph{arXiv preprint
  arXiv:1801.03339}, 2018.

\bibitem{qin2019imperceptible}
Y.~Qin, N.~Carlini, I.~Goodfellow, G.~Cottrell, and C.~Raffel,
  ``{Imperceptible, Robust, and Targeted Adversarial Examples for Automatic
  Speech Recognition},'' \emph{arXiv preprint arXiv:1903.10346}, 2019.

\bibitem{schonherr2018adversarial}
\BIBentryALTinterwordspacing
L.~Sch{\"o}nherr, K.~Kohls, S.~Zeiler, T.~Holz, and D.~Kolossa, ``{Adversarial
  Attacks Against Automatic Speech Recognition Systems via Psychoacoustic
  Hiding}.''\hskip 1em plus 0.5em minus 0.4em\relax The Internet Society, 2019.
  [Online]. Available: \url{https://www.ndss-symposium.org/ndss2019/}
\BIBentrySTDinterwordspacing

\bibitem{abdoli2019universal}
S.~Abdoli, L.~G. Hafemann, J.~Rony, I.~B. Ayed, P.~Cardinal, and A.~L. Koerich,
  ``Universal {A}dversarial {A}udio {P}erturbations,'' \emph{arXiv preprint
  arXiv:1908.03173}, 2019.

\bibitem{yakura2018robust}
H.~Yakura and J.~Sakuma, ``Robust {A}udio {A}dversarial {E}xample for a
  {P}hysical {A}ttack,'' \emph{arXiv preprint arXiv:1810.11793}, 2018.

\bibitem{chen2020devil}
Y.~Chen, X.~Yuan, J.~Zhang, Y.~Zhao, S.~Zhang, K.~Chen, and X.~Wang,
  ``Devil’s whisper: A general approach for physical adversarial attacks
  against commercial black-box speech recognition devices,'' in \emph{29th
  USENIX Security Symposium (USENIX Security 20)}, 2020.

\bibitem{alzantot2018did}
\BIBentryALTinterwordspacing
M.~Alzantot, B.~Balaji, and M.~B. Srivastava, ``{Did you hear that? Adversarial
  Examples Against Automatic Speech Recognition},'' in \emph{Neural Information
  Processing Systems Workshop on Machine Deception 2017}, vol.
  abs/1801.00554.\hskip 1em plus 0.5em minus 0.4em\relax Neural Information
  Processing Systems, 2017. [Online]. Available:
  \url{http://arxiv.org/abs/1801.00554}
\BIBentrySTDinterwordspacing

\bibitem{sugawara2020light}
T.~Sugawara, B.~Cyr, S.~Rampazzi, D.~Genkin, and K.~Fu, ``Light commands:
  laser-based audio injection attacks on voice-controllable systems,''
  \emph{arXiv preprint arXiv:2006.11946}, 2020.

\bibitem{vaidya2015cocaine}
T.~Vaidya, Y.~Zhang, M.~Sherr, and C.~Shields, ``{Cocaine Noodles: Exploiting
  the Gap between Human and Machine Speech Recognition},'' \emph{WOOT},
  vol.~15, pp. 10--11, 2015.

\bibitem{carlini2016hidden}
N.~Carlini, P.~Mishra, T.~Vaidya, Y.~Zhang, M.~Sherr, C.~Shields, D.~Wagner,
  and W.~Zhou, ``{Hidden Voice Commands.}'' in \emph{USENIX Security
  Symposium}, 2016, pp. 513--530.

\bibitem{ds1_tf}
``{DeepSpeech},'' Last accessed in 2019, available at
  \url{https://github.com/mozilla/DeepSpeech}.

\bibitem{wang2019secure}
Y.~Wang, W.~Cai, T.~Gu, W.~Shao, Y.~Li, and Y.~Yu, ``Secure your voice: An oral
  airflow-based continuous liveness detection for voice assistants,''
  \emph{Proceedings of the ACM on Interactive, Mobile, Wearable and Ubiquitous
  Technologies}, vol.~3, no.~4, pp. 1--28, 2019.

\bibitem{wang2019voicepop}
Q.~Wang, X.~Lin, M.~Zhou, Y.~Chen, C.~Wang, Q.~Li, and X.~Luo, ``Voicepop: A
  pop noise based anti-spoofing system for voice authentication on
  smartphones,'' in \emph{IEEE INFOCOM 2019-IEEE Conference on Computer
  Communications}.\hskip 1em plus 0.5em minus 0.4em\relax IEEE, 2019, pp.
  2062--2070.

\bibitem{wang2019defeating}
C.~Wang, S.~A. Anand, J.~Liu, P.~Walker, Y.~Chen, and N.~Saxena, ``Defeating
  hidden audio channel attacks on voice assistants via audio-induced surface
  vibrations,'' in \emph{Proceedings of the 35th Annual Computer Security
  Applications Conference}, 2019, pp. 42--56.

\bibitem{yang2018characterizing}
Z.~Yang, B.~Li, P.-Y. Chen, and D.~Song, ``Characterizing {Audio} {Adversarial}
  {Examples} {Using} {Temporal} {Dependency},'' \emph{arXiv preprint
  arXiv:1809.10875}, 2018.

\bibitem{asr_cost}
``{Apple HomePod build cost hints at thin margins},'' Last accessed in 2020,
  available at
  \url{https://www.slashgear.com/apple-homepod-build-cost-hints-at-thin-margins-14519606/}.

\bibitem{room_size}
``{Average Bedroom Size and Dimensions},'' Last accessed in 2019, available at
  \url{https://www.doorwaysmagazine.com/average-bedroom-size/}.

\bibitem{madry2017towards}
A.~Madry, A.~Makelov, L.~Schmidt, D.~Tsipras, and A.~Vladu, ``Towards {D}eep
  {L}earning {M}odels {R}esistant to {A}dversarial {A}ttacks,'' \emph{arXiv
  preprint arXiv:1706.06083}, 2017.

\bibitem{kurakin2016adversarial}
A.~Kurakin, I.~Goodfellow, and S.~Bengio, ``Adversarial examples in the
  physical world,'' \emph{arXiv preprint arXiv:1607.02533}, 2016.

\bibitem{todisco2019asvspoof}
M.~Todisco, X.~Wang, V.~Vestman, M.~Sahidullah, H.~Delgado, A.~Nautsch,
  J.~Yamagishi, N.~Evans, T.~Kinnunen, and K.~A. Lee, ``Asvspoof 2019: Future
  horizons in spoofed and fake audio detection,'' \emph{arXiv preprint
  arXiv:1904.05441}, 2019.

\bibitem{jooybar2013gpudet}
H.~Jooybar, W.~W. Fung, M.~O'Connor, J.~Devietti, and T.~M. Aamodt, ``{GPUDet}:
  a deterministic {GPU} architecture,'' \emph{ACM SIGARCH Computer Architecture
  News}, vol.~41, no.~1, pp. 1--12, 2013.

\bibitem{kaldi_iflytek}
``{The USTC-iFlytek System For CHiME-4 Challenge},'' Last accessed in 2019,
  available at
  \url{http://spandh.dcs.shef.ac.uk/chime_workshop/chime2016/presentations/CHiME_2016_Du_oral.pdf}.

\bibitem{mozilla_ds_0.4.1}
``{Deep Speech 0.4.1},'' Last accessed in 2019, available at
  \url{https://github.com/mozilla/DeepSpeech/releases/tag/v0.4.1}.

\bibitem{cretu2008casting}
G.~F. Cretu, A.~Stavrou, M.~E. Locasto, S.~J. Stolfo, and A.~D. Keromytis,
  ``Casting {O}ut {D}emons: Sanitizing {T}raining {D}ata for {A}nomaly
  {S}ensors,'' in \emph{2008 IEEE Symposium on Security and Privacy (sp
  2008)}.\hskip 1em plus 0.5em minus 0.4em\relax IEEE, 2008, pp. 81--95.

\bibitem{steinhardt2017certified}
J.~Steinhardt, P.~W.~W. Koh, and P.~S. Liang, ``Certified {D}efenses for {D}ata
  {P}oisoning {A}ttacks,'' in \emph{Advances in neural information processing
  systems}, 2017, pp. 3517--3529.

\bibitem{meng2017magnet}
D.~Meng and H.~Chen, ``{MagNet}: a {T}wo-{P}ronged {D}efense against
  {A}dversarial {E}xamples,'' in \emph{Proceedings of the 2017 ACM SIGSAC
  Conference on Computer and Communications Security}.\hskip 1em plus 0.5em
  minus 0.4em\relax ACM, 2017, pp. 135--147.

\bibitem{jordaney2017transcend}
R.~Jordaney, K.~Sharad, S.~K. Dash, Z.~Wang, D.~Papini, I.~Nouretdinov, and
  L.~Cavallaro, ``Transcend: Detecting {C}oncept {D}rift in {M}alware
  {C}lassification {M}odels,'' in \emph{26th $\{$USENIX$\}$ Security Symposium
  ($\{$USENIX$\}$ Security 17)}, 2017, pp. 625--642.

\bibitem{biggio2015one}
B.~Biggio, I.~Corona, Z.-M. He, P.~P. Chan, G.~Giacinto, D.~S. Yeung, and
  F.~Roli, ``One-and-a-half-class {M}ultiple {C}lassifier {S}ystems for
  {S}ecure {L}earning {A}gainst {E}vasion {A}ttacks at {T}est {T}ime,'' in
  \emph{International Workshop on Multiple Classifier Systems}.\hskip 1em plus
  0.5em minus 0.4em\relax Springer, 2015, pp. 168--180.

\bibitem{bendale2016towards}
A.~Bendale and T.~E. Boult, ``Towards {O}pen {S}et {D}eep {N}etworks,'' in
  \emph{Proceedings of the IEEE conference on computer vision and pattern
  recognition}, 2016, pp. 1563--1572.

\bibitem{xu2017feature}
W.~Xu, D.~Evans, and Y.~Qi, ``Feature {S}queezing: Detecting {A}dversarial
  {E}xamples in {D}eep {N}eural {N}etworks,'' \emph{arXiv preprint
  arXiv:1704.01155}, 2017.

\bibitem{tramer2017ensemble}
F.~Tram{\`e}r, A.~Kurakin, N.~Papernot, D.~Boneh, and P.~McDaniel, ``Ensemble
  {A}dversarial {T}raining: Attacks and {D}efenses,'' \emph{arXiv preprint
  arXiv:1705.07204}, 2017.

\bibitem{lecuyer2019certified}
M.~Lecuyer, V.~Atlidakis, R.~Geambasu, D.~Hsu, and S.~Jana, ``Certified
  {R}obustness to {A}dversarial {E}xamples with {D}ifferential {P}rivacy,'' in
  \emph{2019 IEEE Symposium on Security and Privacy (SP)}.\hskip 1em plus 0.5em
  minus 0.4em\relax IEEE, 2019, pp. 656--672.

\bibitem{select-1}
\BIBentryALTinterwordspacing
D.~J. Plude, J.~T. Enns, and D.~Brodeur, ``The development of selective
  attention: A life-span overview,'' \emph{Acta Psychologica}, vol.~86, no.~2,
  pp. 227 -- 272, 1994. [Online]. Available:
  \url{"http://www.sciencedirect.com/science/article/pii/0001691894900043"}
\BIBentrySTDinterwordspacing

\bibitem{age}
W.~G. on~Speech~Understanding and Aging, ``Speech understanding and aging,''
  \emph{The Journal of the Acoustical Society of America}, vol.~83, no.~3, pp.
  859--895, 1988.

\bibitem{fam}
\BIBentryALTinterwordspacing
F.~Pulvermuller and Y.~Shtyrov, ``Language outside the focus of attention: The
  mismatch negativity as a tool for studying higher cognitive processes,''
  \emph{Progress in Neurobiology}, vol.~79, no.~1, pp. 49 -- 71, 2006.
  [Online]. Available:
  \url{"http://www.sciencedirect.com/science/article/pii/S0301008206000323"}
\BIBentrySTDinterwordspacing

\bibitem{TED}
\BIBentryALTinterwordspacing
C.~Limb, ``Building the musical muscle,'' \emph{TEDMED}, 2011. [Online].
  Available:
  \url{"https://www.ted.com/tals/charles_limb_building_the_musical_muscle#t-367224"}
\BIBentrySTDinterwordspacing

\bibitem{itu2018863}
P.~ITU, ``863 “perceptual objective listening quality prediction”,'' 2018.

\bibitem{schroeder1975models}
M.~R. Schroeder, ``Models of {H}earing,'' \emph{Proceedings of the IEEE},
  vol.~63, no.~9, pp. 1332--1350, 1975.

\bibitem{panayotov2015librispeech}
V.~Panayotov, G.~Chen, D.~Povey, and S.~Khudanpur, ``Librispeech: an {ASR}
  corpus based on public domain audio books,'' in \emph{2015 IEEE International
  Conference on Acoustics, Speech and Signal Processing (ICASSP)}.\hskip 1em
  plus 0.5em minus 0.4em\relax IEEE, 2015, pp. 5206--5210.

\bibitem{kurakin2016adversarialb}
A.~Kurakin, I.~Goodfellow, and S.~Bengio, ``Adversarial machine learning at
  scale,'' \emph{arXiv preprint arXiv:1611.01236}, 2016.

\end{thebibliography}
\section{Acknowledgment}

We would like to thank our reviewers for their insightful comments and suggestions. This work was supported in part by the Defense Advanced Research Projects Agency (DARPA) under agreement number HR0011202008, the National Science Foundation under grant number CNS-1933208. NP acknowledges funding from CIFAR through a Canada CIFAR AI chair, and from NSERC under the Discovery Program and COHESA strategic research network. Any opinions, findings, and conclusions or recommendations expressed in this material are those of the authors and do not necessarily reflect the views of the above listed agencies.

\section{Appendix}\label{appendix}
\subsection{Transferability}\label{transferability_expts}

The transferability property of adversarial samples is a cornerstone of black-box attacks in the image domain (Section~\ref{transferability}). 
In contrast, transferability has had varying success in the audio domain. While signal processing attacks can generate transferable samples, transferability of gradient-based optimization attacks is unclear. While most attack papers do not test for transferability, those that have attempted to, have largely demonstrated unsuccessful results~\cite{cisse2017houdini,yuan2018commandersong}. Considering this major discrepancy between the image and audio domains, we explore experimentally the transferability question for VPSes. Our goal is to ascertain whether adversarial audio samples generated via optimization attacks transfer. 

\begin{table}[ht]
\centering
\begin{tabular}{|l|c|c|c|c|}
\hline
\multirow{2}{*}{\textbf{Target Transcription}} & \multirow{2}{*}{\textbf{\begin{tabular}[c]{@{}c@{}}Benign Samples \\ Perturbed\end{tabular}}} & \multirow{2}{*}{\textbf{\begin{tabular}[c]{@{}c@{}}Adversarial \\ Audio Generated\end{tabular}}} \\
\multicolumn{1}{|l|}{} &  &  \\ \hline
delete my messages & 98 & 1867 \\ \hline 
browse to evil website & 98 & 1086 \\ \hline 
what is the weather today & 98 & 1525 \\ \hline 
go to evil website & 98 & 1652 \\ \hline 
open the door & 98 & 2253 \\ \hline 
transfer money to my account & 98 & 1180 \\ \hline 
the fault dear is not in our stars & 98 & 1312 \\ \hline 
turn off all the cameras & 98 & 1401 \\ \hline 
text mom i need money & 98 & 1400 \\ \hline 
order me some candy & 98 & 1812 \\ \hline 
take a picture & 98 & 1755 \\ \hline 
& \textbf{Total} & 17243 \\ \hline 
\end{tabular}

\caption{The table above shows the results of the transferability experiments for the Carlini et al. attack. The attack was run for 1000 iterations for each of the 98 audio files. There are a varying number of adversarial samples produced for each target transcription. This is because we generated both high confidence (high distortion) and low confidence (low distortion). For example, audio1.wav with target ``open the door'' may be perturbed to generate three high confidence (greater than 0.9) samples at iteration 103, 110, and 200. However, for audio2.wav with target ``text mom I need money'', no high confidence adversarial sample may be possible. Each of the generated adversarial samples (e.g, 2253 for malicious command ``open the door'') was then passed to eight models trained with different initial seeds and one model trained on the same seed. \textit{None of the 17243 samples transfered successfully.}}
\label{tab:transfer_results}
\vspace{-0.8em}
\hrulefill
\vspace{-2.0em}
\end{table}

\subsection{Setup}
\subsubsection{Training}\label{carlini_train}
Testing for transferability requires training a number of ASRs. We use DeepSpeech~\cite{mozilla_ds_0.4.1} for this experiment, even though there are a variety of ASR architectures in use today. This is motivated by two reasons. First, DeepSpeech employs NNs. NN architectures are the most popular and widely used, allowing us to make conclusions applicable to a wider population. Second, as researchers and vendors increasingly phase out non-NN architectures in favor of NNs, our conclusions will be applicable for future systems.

We trained nine DeepSpeech ASRs to achieve the state-of-the-art Word Error Rate (WER) of~8\%~\cite{mozilla_ds_0.4.1}. These ASRs were trained on a cluster of GeForce RTX 2080 Ti GPUs, had 2048 hidden units per layer, were trained on the LibriSpeech dataset~\cite{panayotov2015librispeech} for 13 epochs, and took approximately two days of training each (a total of 180 hours of training time). This training setup closely resembles the official DeepSpeech documentation~\cite{mozilla_ds_0.4.1}. All the ASRs were trained on the same train-test splits and hyper-parameters, except initial random seed. This was the only parameter that was varied, with each ASR being trained on a unique seed value. This is done to emulate the attacker who has perfect knowledge of the ASR's training parameters, except the initial random seed. Next, we trained ASRs with the same hyper-parameters, including the seed. This is the best-case scenario for the attacker as she has absolute knowledge of the ASR's training parameters.

\subsubsection{Adversarial Sample Generation}\label{carlini_sample_gen}

While a number of optimization attacks exist, it is impractical to evaluate every attack with respect to transferability. Luckily, existing optimization attacks do follow the same generic template. These attacks minimize loss functions, use partial derivatives to compute the model's sensitives to the inputs to perturb the input and threshold audible distortion using similar metrics (generally the L2-norm). One such representative attack is Carlini et al.~\cite{carlini2018audio}, which provides the added advantage of being effective against NN based ASRs. Additionally, other optimization attacks have been built directly~\cite{qin2019imperceptible} or indirectly~\cite{yakura2018robust} on this attack. Consequently, we choose the Carlini et al.~\cite{carlini2018audio} to perturb attack audio samples. 

Next, we create a set of audio samples that will be perturbed using the adversarial algorithm. We follow the methodology outlined in Liu et al.~\cite{liu2016delving}. We pass the test set samples from the LibriSpeech dataset to each of our ASR models and pick the 98 samples that all the models transcribed correctly. This methodology is ideal for two reasons. First, by sampling from the test set, we ensure that our audio samples lie within the same distribution as the training data, ensuring consistency. Second, we want our experimental setup to allow for the highest chance of transferability. This enables us to make less error prone conclusions.

We perturbed the set of 98 audio samples that \textit{all} ASRs transcribed correctly. These audio samples were perturbed using 1000 iterations of the attack to force the ASR, called the~\textit{surrogate}, to produce a specific target transcription. We target a total of 11 transcriptions, shown in Table~\ref{tab:transfer_results}. The attacks were used to generate two types of adversarial audio samples: high audible distortion (and high transcription confidence) and low audible noise (and low transcription confidence). This was done specifically to explore whether samples with high audible distortion transfer better than ones with low distortion.

Transferring a targeted transcription is not as easy as we want the target ASR to assign a specific text to the audio. This is because, in the decision space, an adversarially chosen transcription might be very far from the original one. If on the other hand, the chosen transcription is very close the original one, then transferability is more likely. The closest transcription to the original, in the decision space, is the one with the second highest probability after the original. For example, if the original transcription is ``Mary had a little lamb'', the second most likely transcriptions might be ``Mary belittled a lamb''. Intuitively, it will be easier to move from ``Mary had a little lamb''to ``Mary belittled a lamb'' than to ``open the door''. This should be the case as we train the models on the same data partitions and the vocabulary. However, if transferability is not possible in this ideal case, than it is very unlikely for more realistic cases.

To answer whether transferability is possible at all i.e., in the easiest case, we designed the following experiment. We perturbed each audio sample to produce two adversarial samples with the second and third most likely transcriptions as our target attack transcriptions. For example, consider the original audio sample that correctly transcribes to ``Mary had a little lamb''. It's second and third most likely transcriptions are ``Mary belittled a lamb'' and ``Mary had a spittle blam''. We use these as the targets, instead of using a malicious one (``open the door''). Then we repeat the adversarial audio perturbation steps from the previous experiments. 

\subsubsection{Transferring Samples}
Next, the adversarial samples are passed to the ASRs, which we refer to as the~\textit{remote targets}. The transferability is considered a success if the remote target ASRs transcribe the attack audio sample as the attacker's chosen text. We count the number of times transferability succeeded and use it as a metric for transferability success.


\subsection{Results}

\textbf{Are adversarial audio samples perturbed via optimization attacks transferable?}

We first explore this question with regards to ASRs that share all the training parameters, except the random seed. We observed that \textit{none of the adversarial samples successfully transfered} (i.e., none of the remote target ASRs assigned the attacker chosen transcriptions to the audio) (Table~\ref{tab:transfer_results}). This includes both the adversarial samples with high and low distortion. We wanted to explore whether the attack transcriptions existed in the top 10 most probable transcriptions. However, \textit{the attack transcriptions were not present in the top 10 transcriptions}. This experiment demonstrates that transferability, tested over thousands of samples, is unlikely for ASRs trained on different seeds.

\textbf{Do adversarial audio samples transfer if the model is trained on the same seed?}

We trained another set of nine ASRs with the same training set and hyper-parameters, including the random seed. Here, the goal is to check if adversarial samples will transfer between two models that have the exact same training parameters. We used adversarial samples that were generated from the previous experiment. None of the adversarial samples transferred successfully (Table~\ref{tab:transfer_results}). This is a result of the non-determinism introduced in GPUs during training which resulted in ASRs with differing decision boundaries~\cite{jooybar2013gpudet} and contains similar findings in the context of model extraction. Similar to the previous experiment, the \textit{the attack transcriptions were not present in the top 10 transcriptions}. This means even if an attacker has perfect knowledge of the target ASR's training parameters (train-test splits, the hyper-parameters, the architecture, etc),~\textit{adversarial samples still may not transfer unless all sources of non-determinism have been accounted for}.

\textbf{In what cases is transferability possible at all?}

This experiment makes transferability most likely. The chosen attack transcriptions are very close to the original benign ones. We recorded  transferability for this scenario. For the second most likely transcription, the transferability for the low confidence (low distortion) and high confidence (high distortion) was 6.5\% and 40\% respectively. Similarly, the attack transcriptions were in the top 10 labels 68\% of the time. However, for the third most likely transcription, these numbers dropped to 3.0\%, 31\%, and 63\%. 

These numbers reveal two important facets about audio adversarial samples. First, even if the attacker-chosen transcription is almost exactly the same as the original, the probability of a successful transcription is low (6.5\%). This probability drops significantly (from 6.5\% to 3.0\%) when the target transcription moves from second to third most likely. Second, the transferability of low confidence (low distortion) samples is much lower than high confidence (high distortion) samples. This is because high confidence samples are further away from the decision boundaries. As a result, these are more likely to transfer to a different model with an altered decision boundary. This means that~\textit{generally speaking, none of the optimization attacks we experimented with will transfer to other instances of the same ASR.} This is due to substantial non-determinism of the GPU and the community should work towards attacks that can overcome this.

\begin{table}[ht]
\centering
\begin{tabular}{|l|c|c|c|c|}
\hline
\multirow{2}{*}{\textbf{Target Transcription}} & \multirow{2}{*}{\textbf{\begin{tabular}[c]{@{}c@{}}Benign Samples \\ Perturbed\end{tabular}}} & \multirow{2}{*}{\textbf{\begin{tabular}[c]{@{}c@{}}Adversarial \\ Audio Generated\end{tabular}}} \\
\multicolumn{1}{|l|}{} &  &  \\ \hline
delete my messages & 98 & 29968 \\ \hline 
browse to evil website & 98 & 30000 \\ \hline 
what is the weather today & 98 & 30000 \\ \hline 
go to evil website & 98 & 30000 \\ \hline 
open the door & 98 & 64293 \\ \hline 
transfer money to my account & 98 & 30000 \\ \hline 
turn off all the cameras & 98 & 30000 \\ \hline 
order me some candy & 98 & 30000 \\ \hline 
take a picture & 98 & 30000 \\ \hline 
& \textbf{Total} & 239968 \\ \hline 
\end{tabular}

\caption{The table above shows the results of the transferability experiments for the PGD attack. The attack was run for 1000 iterations for each of the 98 audio files. For each attack, we only produced 3000 attack audio files, each of confidence greater than 0.9. Each of the generated adversarial samples was then passed to eight models trained with different initial seeds and one model trained on the same seed. \textit{None of the 239968 samples transfered successfully.}}
\label{tab:transfer_results_pgd}
\vspace{-0.8em}
\hrulefill
\vspace{-2.0em}
\end{table}

\subsection{Transferability (PGD)}\label{transferability_expts_pgd}
To verify whether transferability is hard in ASRs or is a facet of the Carlini attack, we decided to run the same set of experiments with a different optimization attack. This time, we used the Projected Gradient Descent (PGD) method~\cite{kurakin2016adversarialb}, which has been successfully demonstrated against image models. We chose this attack over other optimization attacks as it can produce high confidence attack samples. Most other attacks, including the Carlini attack, stop at the boundary. This is especially true for psychoacoustic attacks~\cite{schonherr2018adversarial,qin2019imperceptible} which are designed to produce imperceptible perturbations instead of high confidence ones. In contrast, we wanted to produce high confidence samples as these are more likely to transfer, as demonstrated in the image space.

\subsubsection{Setup}
We followed the same methodological steps like the ones described for the transferability experiment described in Section~\ref{transferability_expts}. The attack was run for 1000 iterations and clipped using the L infinity norm of 5\%. The clipping was done to ensure that the attack was not completely unconstrained. Otherwise, the audio samples would sound like noise. We generated a total of 250,000 adversarial audio samples over 8 sentences and for each of the 9 models.

\subsubsection{Results}
The results can be seen in Table~\ref{tab:transfer_results_pgd}. None of the 250,000 attack files successfully transferred. This is the case for both the models trained using the same seed and ones trained on different seeds. These results match that of the Carlini attack experiments (Section~\ref{transferability_expts}). This means that the difficulty of transferring attack audio is not a by-product of the attacks themselves. Instead, this stems from some inherent property of ASRs.

\end{document}